\documentclass[11pt]{article}
\usepackage{latexformat}
\pdfoutput=1
\usepackage{amsmath,bbm,array,amsfonts,graphicx,wrapfig,lscape,float,mathtools,multirow,longtable,amsthm,upgreek,mathrsfs}
\usepackage[dvipsnames]{xcolor}
\usepackage{stackrel}
\usepackage[all]{xy}
\usepackage[bottom]{footmisc}
\usepackage{mathrsfs}
\usepackage{tikz}
\usepackage{physics}
\usepackage{bm}
\usepackage[capitalize]{cleveref}
\usepackage{color}
\usepackage{enumerate}
\usetikzlibrary{decorations.pathreplacing}
\usepackage{diagbox}
\numberwithin{equation}{section}
\numberwithin{figure}{section}
\numberwithin{table}{section}
\usepackage{pgfplots}
\pgfplotsset{compat=1.14}
\usepackage{makecell}
\usepackage{lscape}
\usepackage[utf8]{inputenc}
\usepackage{mathtools}
\allowdisplaybreaks
\usepackage[most]{tcolorbox}

\renewcommand{\d}{\textup{d}}

\usepackage[toc,page]{appendix}

\usepackage{tocloft}

\title{Quiver BPS Indices from Crystal Profiles}
\author[a]{Jiakang Bao}
\author[a,b,c]{and Masahito Yamazaki}
	
\affiliation[a]{Graduate School of Physics,
    University of Tokyo, Tokyo 113-0033, Japan}
\affiliation[b]{
	Kavli Institute for the Physics and Mathematics of the Universe,\\
	University of Tokyo, Kashiwa, Chiba 277-8583, Japan}
\affiliation[c]{Trans-Scale Quantum Science Institute, 
    University of Tokyo, Tokyo 113-0033, Japan}
	
\emailAdd{jiakang.bao@phys.s.u-tokyo.ac.jp}
\emailAdd{masahito.yamazaki@ipmu.jp}
	
\preprint{
	\begin{flushright}
		RIKEN-iTHEMS-Report-26
	\end{flushright}
}
	
\abstract{We derive new exact formulae for elliptic genera in two spacetime dimensions with $\mathcal{N}\ge 2$ supercharges, as well as their one- and zero-dimensional counterparts (for Witten indices and matrix-model partition functions). Our results are written as a discrete sum over geometric/combinatorial structures of crystals introduced previously by the authors. The contribution from each crystal may be expressed in terms of the boundary data of a finite substructure of the crystal called the molecules. Our results provide vast generalisations of the celebrated Nekrasov partition function enumerated by the Young diagrams, and uncover new combinatorics underlying the crystal melting. We also analyse the thermodynamic limit of crystals arising from two-dimensional $\mathcal{N}=(0,2)$ theories associated with toric Calabi-Yau fourfolds, where a projection of the Calabi-Yau geometry emerges in the limit.}

\begin{document}
\maketitle

\section{Introduction and Summary}\label{intro}
Supersymmetric gauge theories provide a rich interface between string theory and mathematics. In particular, exact computations of supersymmetric partition functions have revealed deep structures underlying the counting of Bogomol'nyi-Parasad-Sommerfield (BPS) states, enumerative geometry, and representation theory. For supersymmetric gauge theories in two and lower spacetime dimensions, a powerful framework for such computations is given by supersymmetric localisation \cite{Benini:2013nda,Benini:2013xpa,Hori:2014tda,Cordova:2014oxa,Hwang:2014uwa}, where path integrals reduce to finite-dimensional integrals that can be evaluated via the Jeffrey-Kirwan (JK) residues \cite{jeffrey1995localization}.

Following our previous paper \cite{Bao:2025hfu}, we use the JK residue formula to compute the elliptic genera \cite{Witten:1986bf} (Witten indices \cite{Witten:1982df}, resp.~matrix model partition functions) in 2d (1d, resp.~0d) with two ($\mathcal{N}=2$) or four ($\mathcal{N}=4$) supercharges\footnote{For minimally supersymmetric gauge theories, in particular, 2d $\mathcal{N}=(0,1)$ theories, the JK residue formula was further generalised in \cite{Bao:2025xhl}.}. 
It turns out that the combinatorics of these residues admits a suggestive interpretation in terms of discrete geometric/combinatorial structures. For a broad class of quiver gauge theories, the contributions of the poles can be organised nicely in the language of crystals \cite{Bao:2025hfu,Bao:2025dqs}. This in particular includes the toric cases that have been extensively studied over the past two decades \cite{Okounkov:2003sp,Iqbal:2003ds,Ooguri:2009ijd,Ooguri:2009ri,Szendroi:2007nu,Chuang:2009crq,Dimofte:2009bv,Nagao:2009rq,Aganagic:2009cg,Ooguri:2010yk,Cirafici:2010bd,Yamazaki:2010fz,Yamazaki:2011wy,Nekrasov:2017cih,Nekrasov:2018xsb,Nekrasov:2023nai,Galakhov:2023vic,Franco:2023tly,Bao:2024ygr,Szabo:2024lcp,Carcamo:2026yqu}.

From the combinatorial structures of the crystals, one can construct the quiver algebras that encode the BPS spectra \cite{Li:2020rij,Galakhov:2020vyb,Galakhov:2021xum,Galakhov:2021vbo,Noshita:2021ldl,Yamazaki:2022cdg}. They have an elliptic-trignometric-rational hierarchy in line with various partition functions mentioned above. The crystals would then serve as certain representations of these algebras. We also expect the quiver W-algebras \cite{Kimura:2015rgi,Kimura:2016dys,Koroteev:2019byp,Kimura:2023bxy,Kimura:2024xpr,Kimura:2024osv,Kimura:2025lfo,Kimura:2025lig,Nekrasov:2016ydq} from the gauge origami systems to be in the same framework, but capture certain structures from a different facet.

Our main results are presented in Sections \ref{subsec:N4_formula} and \ref{subsec:N2_formula}.
Our partition function is expressed in terms of the boundary data of the molecules, where a molecule $\mathscr{M}$ is a finite substructure of the crystal $\mathscr{C}$ satisfying certain conditions. Schematically, our formula for the grand-canonical partition function reads
\begin{align}
\label{eq:schematic}
    \mathcal{Z}(\bm{q}, \bm{\epsilon}) = \sum_{\mathscr{M} \subset \mathscr{C}} w(\mathscr{M}; \bm{q}, \bm{\epsilon}) \;,
\end{align}
where $\bm{q}$ are the fugacities for the ranks of the gauge groups, 
$\bm{\epsilon}$ are the fugacities included in the definition of the partition function, and $w(\mathscr{M}; \bm{q}, \bm{\epsilon})$
is a weight associated with the molecule $\mathscr{M}$, obtained upon cancellations of the one-loop factors in the integrand of the Jeffrey-Kirwan formula. While the weights are (up to sign) very simple monomials in the exponentiated fugacities in the case of quiver gauge theories associated with toric Calabi-Yau three-folds, in general it is absolutely crucial to have more non-trivial weights in order to properly reproduce the BPS state counting.

For the famous Nekrasov partition function for four-dimensional $\mathcal{N}=2$ theories, the molecules are
enumerated by the (two-dimensional) Young diagrams, and the associated weights $w(\mathscr{M}; \bm{q}, \bm{\epsilon})$ are encoded by the arm and leg lengths of the boxes inside the diagram \cite{Nekrasov:2002qd,Losev:2003py,Nakajima:2003pg,Nekrasov:2003rj}; the formula not only shows the combinatorics of the instanton sum but also illuminates its connections to the Seiberg-Witten geometry and emergent spectral curves. Recently, this expression was recovered in \cite{Jiang:2025luj}, along with the generalisations to higher-dimensional Young diagrams. For more general quivers, the cancellations would be more complicated, as Young diagrams have some special structures not present in more general cases. Nevertheless, the basic idea is similar in spirit (if not in the details), and we could still get some boundary-type formula.

It is still not clear whether these expressions would give rise to something similar to the Seiberg-Witten geometry or how this could reveal deeper insights into the quiver W-algebras and the $qq$-characters. For 2d $\mathcal{N}=(0,2)$ theories associated to toric Calabi-Yau fourfolds with the brane brick models \cite{Franco:2015tna,Franco:2016nwv,Franco:2022iap,Franco:2024lxs}, we find that the molten crystal in the thermodynamic limit may still be described by some smooth limit shape, similar to the brane tilings in \cite{Kenyon:2003uj,Iqbal:2003ds,Ooguri:2009ri}. Therefore, the Calabi-Yau geometry emerges from the discrete data in the sense of \cite{Iqbal:2003ds,Ooguri:2009ri}.

The paper is organised as follows. In \cref{crystal}, we recall the construction of the crystals and the molecules. In \cref{general_N4} (for $\mathcal{N}=4$ theories) and \cref{general_N2} (for $\mathcal{N}=2$ theories), we derive and summarise the expressions of the partition functions in the language of crystal melting, in particular the boundaries of the molecules therein. Some examples are illustrated in \cref{examples}. In \cref{limitshapes}, we discuss the thermodynamic limit of the crystals, where the limit shapes are given by the Ronkin functions. In \cref{JKres}, we briefly review the JK residue formula.

We have made an effort to make the discussions in \cref{crystal}, \cref{general_N4}, and \cref{general_N2}
to be as general as possible (if not the most general), so that they encompass a vast class of quiver gauge theories.
The side effect of this is that the discussions in these sections may look too abstract on a first reading.
In this case, some readers may find it useful to refer to the examples discussed in \cref{examples} when 
reading the general discussions in \cref{crystal}, \cref{general_N4}, and \cref{general_N2}.

\section{Crystals and JK Residues: Review}\label{crystal}

In this section, we quickly summarise the basic ingredients of crystal melting as defined from JK residues \cite{Bao:2025hfu}.

\subsection{Crystals and Molecules}\label{subsec:crystals}

\paragraph{Quiver gauge theory}
We shall consider the quiver gauge theory given by a finite quiver $Q$ with the set of quiver nodes (resp.~arrows) denoted by $Q_0$ (resp.~$Q_1$). For theories with $\mathcal{N}=2$ supersymmetry, there are both chiral multiplets and Fermi multiplets. The set of chiral multiplets and Fermi multiplets would be denoted as $I_c$ and $I_F$, respectively. Moreover, all the quiver nodes would be unitary gauge groups $\text{U}(N_a)$ for all $a\in Q_0$, and there is a framing/flavour node that has a $\text{U}(1)$ symmetry. The arrows would represent the bifundamental/(anti-)fundamental/adjoint matters, and the $F$-term or $J$-/$E$-term relations are polynomial relations involving the fields.

For $\mathcal{N}=2$ theories, we assume that all the Fermi multiplets appear in the $J$-/$E$-term interactions; see \cref{subsec:cancellation_N2} for more discussions on this point. This condition is satisfied for a large class of known 
$\mathcal{N}=2$ theories, e.g.\ those associated with toric Calabi-Yau four-folds.

\paragraph{JK Residue Formula}
As discussed in \cite{Benini:2013xpa,Hori:2014tda,Cordova:2014oxa,Hwang:2014uwa}, the supersymmetric partition function $\mathcal{Z}$ can be computed using the JK residue formula \cite{jeffrey1995localization}:
\begin{equation}
	\mathcal{Z}(\bm{N}, \bm{\epsilon})=\frac{1}{|\mathcal{W}|}\sum_{\bm{u}^*\in\mathfrak{M}^*_\text{sing}}\textrm{JK-Res}_{\bm{u}=\bm{u}^*}(\bm{\mathsf{Q}}(\bm{u}^*),\eta)\, Z_{\textrm{1-loop}}(\bm{\epsilon}, \bm{u}) \;.
\end{equation}
Given a theory with gauge group $G$, whose 
Lie algebra we denote as $\mathfrak{g}$ and Weyl group as  $\mathcal{W}$, the complexified Cartan subalgebra $\mathfrak{h}_{\mathbb{C}}$ of the gauge symmetry is 
parameterised as $\bm{u}=\{u_i\}_{i=1}^N$, with $N$ being the rank of $G$. Similarly, the Cartan subalgebra for the flavour symmetry is parametrised by $\bm{\epsilon}=\{ \epsilon_i \}_{i=1}^F$, with $F$ being the rank of the flavour symmetry group. In practice, the gauge group is given by 
$G=\otimes_{a\in Q_0} \text{U}(N_a)$, where $\bm{N} = (N_a)_{a\in Q_0}$ is the dimension vector; the rank $N$ introduced previously is given by $N=\sum_{a\in Q_0} N_a$.
The one-loop determinant $Z_{\textrm{1-loop}}$ denotes the integrand of the residue, 
and JK-Res denotes the Jeffrey-Kirwan residue. The explanation of the remaining notations can be found in \cref{JKres}. 
Such details, however, are not needed for understanding the rest of the main text.

\paragraph{Crystals} The crystal is an oriented weighted graph $\mathscr{C}=(\mathscr{A},\mathscr{I})$ with the set of vertices $\mathscr{A}$ and the set of arrows $\mathscr{I}$. The vertices (resp.~arrows) would also be referred to as atoms (resp.~chemical bonds). They are given as follows \cite{Bao:2025hfu}.

Given the dimension vector $\bm{N} = (N_a)_{a\in Q_0}$, let us denote by $\mathfrak{U}(\bm{N}, \eta)$ the set of $\bm{u}^*=\left(u^{(a)*}_i \right)$
such that the JK residue at $\bm{u}=\bm{u}^*$ is non-zero and contributes non-trivially (i.e.\ is admissible) under the covector $\eta$. By definition, each
$\bm{u}^* \in \mathfrak{U}(\bm{N}, \eta)$ is an isolated point where at least $\sum\limits_{a \in Q_0} N_a$ hyperplanes meet.
Let $\mathfrak{H}(\bm{u}^*)$ be the set of all hyperplanes meeting at $\bm{u}^*$.

A hyperplane in $\mathfrak{H}(\bm{u}^*)$ takes the form
\begin{equation}
   \left\{u^{(a)}_j-u^{(b)}_i-\dots=0 \right\} \quad (a,b\in Q_0,\quad i\in\{1,\dots,N_a\},\quad j\in\{1,\dots,N_b\} )\;,
\end{equation}
if it is associated with a bifundamental/adjoint matter or a vector multiplet, or 
\begin{equation}
   \left\{u^{(a)}_j-\dots=0 \right\}  \quad (a\in Q_0,\quad i\in\{1,\dots,N_a\}) \;,
\end{equation}
for an (anti-)fundamental matter, where the ellipsis denotes the $u$-independent linear combinations of the residual fugacities $\epsilon_k$. At the intersection of the hyperplanes,
each $u^{(a)*}_i$ is a linear combination of these $\epsilon_k$:
\begin{align}
\label{u_combination}
    u^{(a)*}_i \in\bigoplus_{k=1}^F \mathbb{Z}\, \epsilon_{k} \quad \textrm{for each } a\in Q_0,\quad i\in \{1, \dots N_a\}\;.
\end{align}
Now, all of the $u^{(a)*}_i$ over $i$ and $a$ give
\begin{equation}
    \mathscr{A}(\bm{u}^*) :=\left\{ u^{(a)*}_i  \, \Big| \, a\in Q_0; i=1, \dots, N_a   \right\}
    \subset \bigoplus_{k=1}^F \mathbb{Z}\,\epsilon_k
    \;.
\end{equation}
In other words, the elements of the set $\mathscr{A}(\bm{u}^*)$ are the entries $u^{(a)*}_i$ of the vector $\bm{u}^*$. The set of atoms at level $\bm{N}$ is then
\begin{align}
    \mathscr{A}(\bm{N}, \eta) := \bigcup_{\bm{u}^* \in \mathfrak{U}(\bm{N}, \eta)} \mathscr{A}(\bm{u}^*)\;.
\end{align}
The vertices in the crystal $\mathscr{C}$ is given by the full set of atoms, i.e.,
\begin{align}
    \mathscr{A}(\eta) := \bigcup_{\bm{N} \in \mathbb{Z}_{\ge 0}^{|Q_0|}} \mathscr{A}(\bm{N}, \eta)\;.
\end{align}

The full set of atoms can be written as a direct limit with respect to the partial ordering:\footnote{Note that we have
\begin{align}
    \mathscr{A}(\bm{M}, \eta) \subseteq \mathscr{A}(\bm{N}, \eta) \quad \textrm{when} \quad \bm{M} \leq \bm{N}\;,
\end{align}
where the partial ordering $\bm{M} \leq \bm{N}$ is defined as
\begin{align}
    M_a \leq N_a \quad \textrm{for all} \quad a\in Q_0\;.
\end{align}
This gives an inclusion map $\iota_{\bm{MN}}:\mathscr{A}(\bm{M}, \eta) \hookrightarrow \mathscr{A}(\bm{N}, \eta)$. The direct limit is then $\lim\limits_{\longrightarrow}\mathscr{A}(\bm{N}, \eta)=\bigsqcup\limits_{\bm{N}}\mathscr{A}(\bm{N}, \eta)/\sim$, where the $u\in\mathscr{A}(\bm{M}, \eta)$ and $v\in\mathscr{A}(\bm{N}, \eta)$ are equivalent if and only if there exists some $\bm{L}$ with $\bm{M}\leq\bm{L}$ and $\bm{N}\leq\bm{L}$ such that $\iota_{\bm{ML}}(u)=\iota_{NL}(v)$. Effectively, we may simply think of this limit as taking $N_a\rightarrow\infty$.}
\begin{align}
    \mathscr{A}(\eta) = \lim_{\longrightarrow} \mathscr{A}(\bm{N}, \eta) \;.
\end{align}

Let us denote the atoms using $\mathfrak{a}$ associated to the node $(a)$ in the quiver. Suppose that we have two atoms $\mathfrak{a},\mathfrak{b}\in\mathscr{A}$. We draw an arrow $I$ from $\mathfrak{b}$ to $\mathfrak{a}$ if the following two conditions are satisfied:
\begin{itemize}
    \item There exists $\bm{N} \in \mathbb{Z}_{\ge 0}^{|Q_0|}$ and $\bm{u}^{*} \in \mathfrak{U}(\bm{N}, \eta)$
        such that $\mathfrak{a}, \mathfrak{b} \in \mathscr{A}(\bm{u}^*)$.
    \item There exists a hyperplane of the form
    \begin{align}
        \left\{ u^{(a)}_j-u^{(b)}_i-\dots=0 \right\}
    \end{align}
    inside $\mathfrak{H}(\bm{u}^*)$.
\end{itemize}
This gives the set $\mathscr{I}$ of chemical bonds.

\paragraph{The no-overlap condition} We shall further assume the no-overlap condition, which will play a significant role throughout the rest of this paper. Suppose that we are given an atom $\mathfrak{a}$ in $\mathscr{A}(\eta)$
such that $\mathfrak{a} \in \mathscr{A}(\bm{N}, \eta)$
for some $\bm{N}$. In general, this can have multiple realisations inside $\mathscr{A}(\bm{N}, \eta)$, so that
\begin{align}
    u_i^{(a)*}  =u_j^{(b)*}
\end{align}
for a different pair $(i, a), (j, b)$ (in other words, 
either $i\ne j$, or $a\ne b$ if $i=j$), but with the same $\bm{u}^* \in \mathscr{A}(\bm{N}, \eta)$. If this ever happens for some $\bm{N}$, we say that the no-overlap condition is violated; otherwise, we say that the no-overlap condition is satisfied.

This is equivalent to the condition that any $\mathscr{A}(\bm{u}^*)$ contains
$\sum\limits_{a\in Q_0} N_a$ different elements for any $\bm{u}^* \in \mathfrak{U}(\bm{N}, \eta)$ for any dimension vector. Intuitively, this means that the atoms are not allowed to overlap in the crystal.

When the no-overlap condition is satisfied, for an atom $\mathfrak{a}$ there exists a unqiue vertex $a\in Q_0$ such that $\mathfrak{a}$ can be written as $u_i^{(a)*}$ for some $i$; we may call this vertex $a$ the associated vertex of the atom $\mathfrak{a}$. To simplify the presentations, in the following we use the simplified notation where the atoms $\mathfrak{a}, \mathfrak{b}, \mathfrak{c}, \dots$ 
have associated vertices $a, b, c$.\footnote{More pricesely, we can define 
a function $v: \mathscr{A}\to Q_0$ by assigning the associated vertex $a=v(\mathfrak{a})$ to the atom $\mathfrak{a}$. In our convention, we implicitly assume $a =v (\mathfrak{a})$, $b =v (\mathfrak{b})$, $c =v (\mathfrak{c})$.
This notation helps to avoid clutter in notations in what follows.}

\paragraph{Comments on the Definitions of Crystals}

In our discussion, following \cite{Bao:2025hfu} we have defined the crystal in such a way that an atom is a point on the lattice of flavour fugacities.
This is different from the definitions in earlier literature on crystals, e.g.\ in \cite{Ooguri:2009ijd,Li:2020rij},
where an atom is defined to be an element of the path algebra associated with a quiver with relations. While one should in general be careful in distinguishing between the two different contexts, 
we hope no confusion arises here, since strictly speaking we only need the definition of crystals in \cite{Bao:2025hfu}. 
We find it intuitive, however, to occasionally use the notation 
motivated by the other definition, such as 
to denote the flavour charge of an atom $\mathfrak{a}$
as $\epsilon(\mathfrak{a})$, despite the fact that in our notation
$\mathfrak{a}$ is in itself given by the flavour charge.
While some readers may prefer to disregard the notation $\epsilon(\mathfrak{a})$ and simply read it as $\mathfrak{a}$, our intention is that this notation makes it clearer that we are discussing flavour weights.

\paragraph{Molecules} Given the crystal $\mathscr{C}=(\mathscr{A}, \mathscr{I})$, we can define the molecule
as a finite subset $\mathscr{M} \subset \mathscr{A}$
such that the following condition (melting rule) is satisfied.\footnote{We occassionally write $\mathscr{M} \subset \mathscr{A}$ as $\mathscr{M} \subset \mathscr{C}$ by abuse of notation.} Suppose that $\mathfrak{a}, \mathfrak{b} \in \mathscr{A}$ with an arrow $I \in \mathscr{I}$ connecting from $\mathfrak{a}$ to $\mathfrak{b}$. If $\mathfrak{b} \in \mathscr{M}$, then $\mathfrak{a} \in \mathscr{M}$. The complement $\mathscr{A}\backslash\mathscr{M}$ would be referred to as the molten crystal.

In the molecule $\mathscr{M}$, the atoms would be connected by the chemical bonds as discussed above. As a graph, $\mathscr{M}$ may not be connected. We shall refer to each connected component as a sub-molecule.

\subsection{Framing and Flavour Weights}

\paragraph{Framing}

In this paper, we consider a framed quiver, 
where an original quiver (called the unframed quiver) is extended an extra vertex $\infty$ (called the framing vertex), as well as edges connecting one of the framing vertices to the nodes of the unframed quiver. 
Such arrows connecting a framing vertex
represent fields in the fundamental or anti-fundamental representation of the product gauge group; in contrast, other fields associated with the edges of the unframed quiver are in 
either a bifundamnetal or an adjoint representation of the product gauge group. See \cite{Galakhov:2021xum} for more details on framings.

\paragraph{Cyclic chambers} 

While the JK residue formula and the crystal melting description work for any generic value of $\eta$, the partition functions (in 1d and 0d) depend on the choice of $\eta$, representing the wall-crossing phenomena \cite{Hori:2014tda,Cordova:2014oxa}. 

In this paper, we shall only consider the special value $\eta=(1,1,\dots,1)$, and often drop $\eta$ from the rest of the notations.\footnote{It is clear that the combinatorics discussed in this paper generalizes to a more general choice of $\eta$. We will, however, defer a detailed analysis for future work.}
This special choice of $\eta$ leads to some simplifications.
First, the growth of the crystal always starts with an arrow starting with a framing node.
In this case, an atom in the crystal is associated with an equivalence class of open paths starting at the framing vertex.
Second, a single-step growth of the crystal is
always described by the addition of one and only atom. This is in contrast to a general choice of $\eta$, where multiple atoms may be added to the molecule in a single step \cite{Galakhov:2024foa,Bao:2025hfu}. To describe the third simplification, we first need to comment on flavour fugacities.

\paragraph{Flavour fugacities}

For a chiral multiplet $\chi$ and a Fermi multiplet $\Lambda$,
we will denote the associated flavour fugacities by $\epsilon_\chi$ and $\epsilon_\Lambda$, respectively.

These parameters are not independent inside the Lagrangian, and 
can be parameterized by a set of 
linearly-independent flavour fugacities $\epsilon_i$ ($i=1, \dots, F$).
While in general there is no canonical choice of such a basis, 
we can choose a basis such that all the $\epsilon_{\chi}$, $\epsilon_{\Lambda}$
take values in the lattice $\Delta$ of integer-coefficient linear combinations of the basis weights $\epsilon_i$ ($i=1, \dots, F$).
Since an atom in the crystal is represented by a path starting with a framing vertex and following a collection of arrows describing chiral multiplets $\chi$, we  can write the flavour weight of an atom as 
\begin{equation}
    \epsilon(\mathfrak{a})
\in 
    \Delta:=\bigsqcup_{l\in \mathscr{O}} \left(v_l +  \bigoplus_{k=1}^F \mathbb{Z}\, \epsilon_{k}  \right)\;.
 \end{equation}
Here $v_l$ represents the flavour weight of a fundamental matter originating from the $l^\text{th}$ framing vertex.
When $\epsilon(\mathfrak{a})=v_l$, the atom $\mathfrak{a}$ is represented by a single arrow starting from a framing vertex, and we say that the atom $\mathfrak{a}$ is an initial atom. The set of initial atoms is denoted as $\mathscr{O}$.

The crystal as well as the flavour weights have a particular positivity property for a special choice of $\eta=(1, 1, \dots, 1)$.
In this case, the crystal sits inside a positive cone $\Delta^{+}$
in the space of the flavour fugacities $\epsilon_k$ ($k=1, \dots, F)$; in other words, there exists a choice of basis of flavour fugacities such that the crystal is contained in the positive cone 
\begin{align}
    \Delta^{+}:=\bigsqcup_{l\in \mathscr{O}} \left(v_l +  \bigoplus_{k=1}^F \mathbb{Z}_{\ge 0}\, \epsilon_{k}  \right)\;.
\end{align}
In practice, this means that a chiral multiplet $\chi$
has a flavour parameter $\epsilon_{\chi}$ such that 
$\epsilon_{\chi} \in \Delta^{+}$, while a 
Fermi multiplet $\Lambda$ enforcing an either a $J$-term or $E$-term constraint has a flavour parameter $\epsilon_{\Lambda}$
such that $-\epsilon_{\Lambda} \in \Delta^{+}$.\footnote{
If the $J$/$E$-term involves chiral multiplets $\chi_1, \chi_2, \dots$, when the associated Fermi multiplet $\Lambda$ has the flavour parameter $\epsilon_\Lambda=-\epsilon_{\chi_1}-\epsilon_{\chi_2}-\dots$.}
This means that a chiral (Fermi) multiplet $\chi$ ($\Lambda$)
goes ``forward'' (``backward'') in the crystal.
 
\section{General Formula for \texorpdfstring{$\mathcal{N}=4$}{N=4} Crystals}\label{general_N4}

In this section, we derive our formula
for the grand-canonical partition function for 
general $\mathcal{N}=4$ theories.
One feature of $\mathcal{N}=4$ theory, compared with a general $\mathcal{N}=2$ theory,
is the existence of an extra $\text{U}(1)$ R-symmetry, which looks like an extra flavour symmetry
when regarded as an $\mathcal{N}=2$ theory. We denote by $\varepsilon$
the corresponding flavour parameter,
and in the notations distinguish it from the rest of the flavour parameters that are collectively denoted by $\bm{\epsilon}$. The parameter $\varepsilon$ will play an extremely crucial role in what follows.

\subsection{JK Integrand} 

For the analysis of the weights associated with a crystal, we need formulae for the integrands of the JK residue, which arise as one-loop determinants in the supersymmetric localization of the path integral.

Throughout this paper, we use the function defined as
\begin{equation}
	\zeta(z)=\begin{cases}
		\displaystyle\frac{\text{i}\theta_1(\tau,z)}{\eta(\tau)}\;, &\text{elliptic (2d)}\;,\\
		2\text{i}\sin(\pi z)\;, 
        &\text{trigonometric (1d)}\;,\\
		z \;,&\text{rational (0d)}\;,
	\end{cases}\label{zeta}
\end{equation}
where
\begin{equation}
	\eta(\tau)=\mathfrak{q}^{1/24}\prod_{k=1}^{\infty}\left(1-\mathfrak{q}^k\right)\;,
    \quad
    \theta_1(\tau,z)=-\text{i}\mathfrak{q}^{1/8}y^{1/2}\prod_{k=1}\left(1-\mathfrak{q}^k\right)\left(1-y\mathfrak{q}^k\right)\left(1-y^{-1}\mathfrak{q}^{k-1}\right)\;,
\end{equation}
with $\mathfrak{q}=\text{e}^{2\pi\text{i}\tau}$ and $y=\text{e}^{2\pi\text{i}z}$.

The integrand $Z_\text{1-loop}$ factorises into the contributions from the vector/chiral multiplet contributions:
\begin{equation}
	Z_{\text{1-loop}}(\bm{u}) = \prod_VZ_V( \varepsilon,\bm{u})\prod_{\chi}Z_{\chi}(\bm{\epsilon}, \varepsilon, \bm{u})\;.
\end{equation}
\begin{itemize}
	\item Vector multiplet:
	\begin{equation}
		Z_V(\varepsilon, \bm{u})=[\text{d}\bm{u}] \, \xi (N)\left(-\frac{1}{\zeta(\varepsilon)}\right)^N\prod_{i\neq j}^N\frac{-\zeta(u_i-u_j)}{\zeta(u_i-u_j+\varepsilon)} \;.
	\end{equation}
	where we defined
\begin{equation}
\label{xi_def}
	\xi(N):=\begin{cases}
		\eta(\tau)^{2N}\;,&\text{elliptic (2d)}\;,\\
		1\;,&\text{trigonometric (1d), rational (0d)}\;,
	\end{cases}
\end{equation}
and
\begin{equation} \label{du_def}
	[\text{d}\bm{u}]:=\begin{cases}
		\displaystyle
		\prod\limits_{i=1}^N\text{d}(2\pi\text{i}u_i)\;,&\text{elliptic (2d), trigonometric (1d)}\;,\\
		\displaystyle
		\prod\limits_{i=1}^N \text{d}u_i\;,&\text{rational (0d)}\;.
	\end{cases}
\end{equation}
	
	\item Chiral multiplet with $\text{U}(1)$ symmetry charge $\epsilon_{\chi}$:
	\begin{equation}
    \label{chiral_N4_determinant}
		Z_{\chi}(\bm{\epsilon}, \varepsilon, \bm{u})=
			\displaystyle
			\prod\limits_{i=1}^{N_s}\prod\limits_{j=1}^{N_t}\frac{-\zeta\left(u^{(t)}_j-u^{(s)}_i+\varepsilon-\epsilon_{\chi}\right)}{\zeta\left(u^{(t)}_j-u^{(s)}_i-\epsilon_{\chi}\right)}\;.
	\end{equation}
\end{itemize}
Here, $s$ and $t$ indicate that this is a (bi)fundamental from node $(s)$ to node $(t)$, and can be understood as an adjoint when $s=t$.

\subsection{Cancellation Mechanisms}\label{subsec:cancellation_N4}
To write down the expression of the partition function for a given crystal, let us focus on a fixed atom, say corresponding to the variable $u^{(a)}_i$ in the one-loop determinant. 

The factors involving $u^{(a)}_i$ are the followings:
\begin{itemize}
    \item From the vector multiplet:
    \begin{equation}
        \frac{-\color{blue}\zeta\left(u^{(a)}_i-u^{(a)}_j\right)}{\color{red}\zeta\left(u^{(a)}_i-u^{(a)}_j+\varepsilon\right)}
    \end{equation}
    for any $j\neq i$.
    \item From the chiral multiplets without the framing:
    \begin{equation}
        \prod_{\chi\in\{b\rightarrow a\}}\frac{-\color{red}\zeta\left(u^{(a)}_i-u^{(b)}_j+\varepsilon-\epsilon_\chi\right)}{\color{blue}\zeta\left(u^{(a)}_i-u^{(b)}_j-\epsilon_\chi\right)}\;,
    \end{equation}
    where for the adjoint chiral multiplets, we have omitted the constant prefactor independent of $u$. Notice that we only need to consider the arrows ending on the node $a$ here, namely factors of the form $\zeta\left(u^{(a)}_i-u^{(b)}_j+\dots\right)$. For the factors of the form $\zeta\left(u^{(c)}_k-u^{(a)}_i+\dots\right)$ coming from the arrows emanating from $a$, they are considered when the fixed atom corresponds to $u^{(c)}_k$.
    \item From the chiral multiplets with the framing:
    \begin{equation}
        \prod_{l=1}^{\#[\infty\rightarrow a]}\frac{-\zeta\left(u^{(a)}_i-v_l+\varepsilon\right)}{\zeta\left(u^{(a)}_i-v_l\right)}\quad\text{and}\quad\prod_{l=1}^{\#[a\rightarrow\infty]}\frac{-\zeta\left(v_l-u^{(a)}_i+\varepsilon\right)}{\zeta\left(v_l-u^{(a)}_i\right)}\;.
    \end{equation}
\end{itemize}
Recall that we have assumed generic parameters $v_k$, so there would be no cancellations among the factors from the framing at this stage. For the other factors, when ${\color{red}u^{(b)}_k}={\color{blue}u^{(b)}_j}+\varepsilon$, the blue factors with ${\color{blue}\left(u^{(a)}_i,u^{(b)}_j\right)}$ would get cancelled by the red factors with ${\color{red}\left(u^{(a)}_i,u^{(b)}_k\right)}$. For instance, ${\color{blue}\zeta\left(u^{(a)}_i-u^{(a)}_j\right)}$ and ${\color{red}\zeta\left(u^{(a)}_i-u^{(a)}_k+\varepsilon\right)}$ would cancel each other in the contributions from the vector multiplet, and likewise for the chiral multiplets.

This means that most of the factors actually make a pair between numerators and denominators,
and hence cancel out from the final expression. 
Some of these factors, however, do not have any cancelling pairs
and would survive into the final expression.
Whether or not this happens depends on the positions of the atoms.
\begin{itemize}
    \item There would be no $u^{(b)}_k$ when $\mathfrak{b}_j\in\partial_+\mathscr{M}$, which is defined as
\begin{equation}
    \partial_{+} \mathscr{M}
    := 
    \left\{
    \mathfrak{a} \in \mathscr{M} \,|\, \epsilon(\mathfrak{a})+\varepsilon\notin\mathscr{M}
    \right\}\;.
\end{equation}
The blue factors would remain in this case.

    \item There would be no $u^{(b)}_j$ when $\mathfrak{b}_k\in\partial_-\mathscr{M}$, which is defined as
    \begin{equation}
    \partial_{-} \mathscr{M}
    := 
    \left\{
    \mathfrak{a} \in \mathscr{M} \,|\, \epsilon(\mathfrak{a})-\varepsilon\notin\mathscr{M}
    \right\}\;.
\end{equation}
The red factors would remain in this case.
\end{itemize}
Altogether, these residual contributions to the BPS partition functions are associated with the boundary $\partial_+\mathscr{M}\cup\partial_-\mathscr{M}$ of the molecule.

In addition to the factors above, one needs to take into account the factors originating from vector multiplets.
Recall that for the vector multiplets and the adjoint chiral multiplets, we require $i\neq j$. 
Therefore, the factors\footnote{Notice that besides $u^{(a)}_i$, the other variable is $u^{(a)}_j=u^{(a)}_i-\varepsilon$ (resp.~$u^{(a)}_j=u^{(a)}_i+\varepsilon$) in the first (resp.~second) factor.}
    \begin{equation}
        -\zeta\left(u^{(a)}_i-u^{(a)}_i+\varepsilon\right)\;,\quad\frac{1}{\zeta\left(u^{(a)}_i-u^{(a)}_i-\varepsilon+\varepsilon\right)}
    \end{equation}
    from the vector multiplet would remain uncancelled. 
Nevertheless, we may safely discard them due to the following reasons. The first factor would be cancelled by a prefactor $-1/\zeta(\varepsilon)$ in $\xi$, and there are $|\mathscr{M}\backslash\partial_-\mathscr{M}|$ of such factors\footnote{The remaining $|\partial_- \mathscr{M}|$ such prefactors, $-1/\zeta(\varepsilon)$, in $\xi$ would be cancelled by the contributions from the chiral multiplets.}. For the second factor, it would contribute as a pole in the JK residue formula, and this pole corresponds to the atom with $u^{(a)}_i+\varepsilon$ in the molecule. This indicates that we further need to remove the poles corresponding to the atoms in $\partial_-\mathscr{M}$ in the contribution from the chiral multiplets in the final expression of the partition function. Likewise, the factors
\begin{equation}
    \frac{1}{\zeta\left(u^{(a)}_i-u^{(a)}_i+\varepsilon-\epsilon_\chi\right)}\;,\quad-\zeta\left(u^{(a)}_i-u^{(a)}_i-\varepsilon+\varepsilon-\epsilon_\chi\right)\;,
\end{equation}
namely, the constant prefactors from the chiral multiplets, would remain uncancelled.

Now, let us consider the factors from the framing. When $u^{(a)}_j=u^{(a)}_i+\epsilon$, the factors $\zeta\left(u^{(a)}_i-v_l+\varepsilon\right)$ in the numerator and $\zeta\left(u^{(a)}_j-v_l\right)$ in the denominator would get cancelled. This is likewise for $\zeta\left(v_l-u^{(a)}_i\right)$ and $\zeta\left(v_l-u^{(a)}_j+\varepsilon\right)$. Again, there are two situations with some factors surviving after the cancellations as above\footnote{We also need to discard the factors corresponding to the initial atoms in the expression of the partition function since they contribute as poles in the integral.}.

\subsection{The \texorpdfstring{$\mathcal{N}=4$}{N=4} Formula}\label{subsec:N4_formula}

We are now ready to write down our formula.
We will write down the partition grand-canonical ensemble, assigning a weight 
\begin{equation}
    \bm{q}^{\bm{N}}=\prod_{a=1}^nq_a^{N_a}
\end{equation}
to each quiver gauge theory with dimension vector $\bm{N}$:
\begin{align}
    \mathcal{Z}(\bm{q}, \bm{\epsilon}, \varepsilon)
    = \sum_{\bm{N}} \bm{q}^{\bm{N}}
    \mathcal{Z}(\bm{N}, \bm{\epsilon}, \varepsilon) \;.
\end{align}
In the language of molecules, an entry $N_a$ in the dimension vector is the number of atoms of type $a$ in the molecule; by definition, they sum up to the total number of atoms in the molecule $\sum\limits_aN_a=N=|\mathscr{M}|$.

We need to take into account 
one-loop determinants. Since the choice of the molecule already takes care of the poles and zeros, we can substitute the values of the $u$'s, as determined by an atom $\mathfrak{a}$ in the molecule $\mathscr{M}$, into the one-loop determinants.
To explain the result, it is useful 
to introduce the following notation:
for two atoms $\mathfrak{a}_i$ and $\mathfrak{b}_j$, we write
\begin{equation}
\label{def_E_ij}
    E_{ij}(x) :=\zeta(\epsilon(\mathfrak{a}_i)-\epsilon(\mathfrak{b}_j)+x), \quad E_{00}(x)=\zeta(x)\;.
\end{equation}
Notice that $E_{ij}(x)=-E_{ji}(-x)$ and $E_{jj}(x)=\zeta(x)$.
In the notation \eqref{def_E_ij}, we have chosen to label the atoms by indices $i, j, \dots$, as in $\mathfrak{a}_i, \mathfrak{b}_j$.
This is a redundant notation, and one may prefer the notation
$E_{\mathfrak{a}, \mathfrak{b}}(x) =\zeta(\epsilon(\mathfrak{a})-\epsilon(\mathfrak{b})+x)$, or 
$E_{\mathfrak{a}_i, \mathfrak{b}_j}(x) =\zeta(\epsilon(\mathfrak{a}_i)-\epsilon(\mathfrak{b}_j)+x)$.
We will, however, find it useful to use the notation $E_{ij}(x)$ for notational simplicity.

Our final formula for the $\mathcal{N}=4$ partition function reads
\begin{tcolorbox}[ams align]
\begin{aligned}
    \mathcal{Z}(\bm{q}, \bm{\epsilon}, \varepsilon)=\sum_{\mathscr{M}\subset \mathscr{C}}\bm{q}^{\bm{N}}\xi(\varepsilon, N) \, \mathcal{Z}_V(\varepsilon)\, \mathcal{Z}_\chi(\bm{\epsilon}, \varepsilon)\, \mathcal{Z}_\mathfrak{f}(\bm{\epsilon}, \varepsilon)\;,
\end{aligned}
\end{tcolorbox}
The factors $\mathcal{Z}_V\, \mathcal{Z}_\chi\, \mathcal{Z}_\mathfrak{f}$ contributing to $\mathcal{Z}(\bm{q}, \bm{\epsilon})$ after the process of cancellations and the integration are
\begin{subequations}
\begin{tcolorbox}[ams align]
    &\mathcal{Z}_V(\varepsilon)=\prod_{a\in Q_0}\frac{\prod\limits_{\substack{\mathfrak{a}_i\in\mathscr{M}, \mathfrak{a}_j\in\partial_+\mathscr{M}\\ \mathfrak{a}_j\ne \mathfrak{a}_i}}E_{ij}(0)}{\prod\limits_{\substack{\mathfrak{a}_i\in\mathscr{M}, \mathfrak{a}_j\in\partial_-\mathscr{M}\\
    \mathfrak{a}_j\ne \mathfrak{a}_i}}E_{ij}(\varepsilon)}\;,\\
    &\mathcal{Z}_\chi(\bm{\epsilon}, \varepsilon)=(-1)^{\sum\limits_{a,b\in Q_0}\#[b\rightarrow a]N_aN_b}
    \prod_{a,b\in Q_0}
    \prod_{\chi\in\{b\rightarrow a\}}
    \frac{\prod\limits_{\substack{\mathfrak{a}_i\in\mathscr{M},\mathfrak{b}_j\in\partial_-\mathscr{M}\\ \epsilon(\mathfrak{b}_j)\neq\epsilon(\mathfrak{a}_i)-\epsilon_\chi}}E_{ij}(\varepsilon-\epsilon_\chi)}{\prod\limits_{\substack{\mathfrak{a}_i\in\mathscr{M},\mathfrak{b}_j\in\partial_+\mathscr{M}\\ \epsilon(\mathfrak{b}_j)\neq\epsilon(\mathfrak{a}_i)-\epsilon_\chi}}E_{ij}(-\epsilon_\chi)}\;,\\
    &\mathcal{Z}_\mathfrak{f}(\bm{\epsilon}, \varepsilon)
    =(-1)^{\sum\limits_{a\in Q_0}N_a(\#[\infty\rightarrow a]+\#[a\rightarrow\infty])}\left(\prod_{a\in Q_0}\left(\prod_{\chi\in\{\infty\rightarrow a\}}\frac{\prod\limits_{\mathfrak{a}_i\in\partial_+\mathscr{M}}E_{00}(\epsilon(\mathfrak{a}_i)-v_\chi+\varepsilon)}{\prod\limits_{\mathfrak{a}_i\in\partial_-\mathscr{M}\backslash\mathscr{O}}E_{00}(\epsilon(\mathfrak{a}_i)-v_\chi)}\right)\right.\nonumber\\
    &\qquad\qquad\left.\left(\prod_{\chi\in\{a\rightarrow\infty\}}\frac{\prod\limits_{\mathfrak{a}_i\in\partial_-\mathscr{M}}E_{00}(v_\chi-\epsilon(\mathfrak{a}_i)+\varepsilon)}{\prod\limits_{\mathfrak{a}_i\in\partial_+\mathscr{M}}E_{00}(v_\chi-\epsilon(\mathfrak{a}_i))}\right)\right)\;.
\end{tcolorbox}
\end{subequations}
In these expressions, we encounter conditions in the product---for example, we have the condition  $\epsilon(\mathfrak{b}_j)\neq\epsilon(\mathfrak{a}_i)-\epsilon_\chi$ in $\mathcal{Z}_\mathfrak{\chi}$, since if this condition is not satisfied, the factor $E_{ij}(\varepsilon-\epsilon_{\chi})$ gives zero. Similarly, we need to dispose of the factors corresponding to zeros and poles of the one-loop determinants, since these are already dealt with when we identified the molecules of the crystal as combinatorial descriptions of the non-vanishing contributions.
We also note that we have included signs in the expressions for $\mathcal{Z}_{\chi}$ and $\mathcal{Z}_\mathfrak{f}$ 
that originate from the signs in the one-loop determinant \eqref{chiral_N4_determinant}. Note that in $\mathcal{Z}_V$, the corresponding sign is given by $(-1)^{N_a(N_a-1)}=1$, and hence is trivial for each $a$.

\section{General Formula for \texorpdfstring{$\mathcal{N}=2$}{N=2} Crystals}\label{general_N2}

We now discuss the case of $\mathcal{N}=2$. Notice that we will often use the same notation as already introduced in the 
previous section, e.g., the function $\zeta$ as defined in \eqref{zeta}. We will, however, find important differences from the $\mathcal{N}=4$ counterparts, both in the structure of the cancellation mechanisms and definitions of crystal boundaries.

\subsection{JK Integrand} 

The integrand $Z_\text{1-loop}$ factorises into the contributions from the vector/chiral/Fermi multiplet contributions:
\begin{align}
	Z_{\text{1-loop}}(\bm{\epsilon}, \bm{u}) = \prod_VZ_V( \bm{u})\prod_{\chi}Z_{\chi}(\bm{\epsilon} , \bm{u})\prod_{\Lambda}Z_{\Lambda}(\bm{\epsilon} , \bm{u})\;,
\end{align}
where we have the following:
\begin{itemize}
	\item Vector multiplet:
	\begin{equation}
		Z_V(\bm{u})=[\text{d}\bm{u}] \, \xi(N) \prod_{i\neq j}^N(-\zeta(u_i-u_j)) \;.
	\end{equation}
	(See \eqref{xi_def} and \eqref{du_def}
    for the definition of $\xi(N)$ and $[\text{d}\bm{u}]$, respectively.)

	\item Chiral multiplet:
	\begin{equation}
		Z_{\chi}(\bm{\epsilon}, \bm{u})=
			\displaystyle
			\prod\limits_{i=1}^{N_s}\prod\limits_{j=1}^{N_t}\frac{1}{\zeta\left(u^{(t)}_j-u^{(s)}_i-\epsilon_{\chi}\right)}\;.
	\end{equation}
	\item Fermi multiplet:
	\begin{equation}
		Z_{\Lambda}(\bm{\epsilon}, \bm{u})=
			\displaystyle
			\prod\limits_{i=1}^{N_s}\prod\limits_{j=1}^{N_t}\left(-\zeta\left(u^{(t)}_j-u^{(s)}_i-\epsilon_{\Lambda}\right)\right)\;.
	\end{equation}
\end{itemize}

\subsection{Cancellations Mechanisms}\label{subsec:cancellation_N2}
Let us now analyse the cancellations of the numerators and denominators. As in the $\mathcal{N}=2$ case, we first fix an atom labelled by $u^{(a)}_i$. The factors involving $u^{(a)}_i$ are the followings:
\begin{itemize}
    \item From the vector multiplet:
    \begin{equation}
        -\zeta\left(u^{(a)}_j-u^{(a)}_i\right)
    \end{equation}
    for any $j\neq i$.
    \item From the chiral multiplets without the framing:
    \begin{equation}
        \prod_{\chi\in\{b\rightarrow a\}}\frac{1}{\zeta\left(u^{(a)}_i-u^{(b)}_j-\epsilon_\chi\right)}\;,
    \end{equation}
    where for the adjoint chiral multiplets, we have omitted the constant prefactor independent of $u$.
    \item From the Fermi multiplets without the framing:
    \begin{equation}
        \prod_{\Lambda\in\{a\rightarrow b\}}\left(-\zeta\left(u^{(b)}_j-u^{(a)}_i-\epsilon_\Lambda\right)\right)\;,
    \end{equation}
    where we have again omitted the constant prefactor independent of $u$ for the adjoint Fermi multiplets.
    \item From the matter multiplets with the framing:
    \begin{equation}
        \frac{\prod\limits_{\Lambda\in\{\infty\rightarrow a\}}\left(-\zeta\left(u^{(a)}_i-v_\Lambda\right)\right)}{\prod\limits_{\chi\in\{\infty\rightarrow a\}}\zeta\left(u^{(a)}_i-v_\chi\right)}\quad\text{and}\quad\frac{\prod\limits_{\Lambda\in\{a\rightarrow\infty\}}\left(-\zeta\left(v_\Lambda-u^{(a)}_i\right)\right)}{\prod\limits_{\chi\in\{a\rightarrow\infty\}}\zeta\left(v_\chi-u^{(a)}_i\right)}\;.
    \end{equation}
\end{itemize}

In contrast to the $\mathcal{N}=4$ cases, some of the multiplets only contribute to the numerator while some only contribute to the denominator. Therefore, the cancellations are among the factors from different multiplets.

As already stated in \cref{subsec:crystals}, we assume in this paper that all the Fermi multiplets appear in the $J$-/$E$-term interactions. In other words, there would be no extra Fermi multiplets ``floating around'' with generic weights which do not cancel anything at any place\footnote{If there are such Fermi multiplets, the partition function would simply have some extra factors from the contributions of these Fermi multiplets.}. Now, for the multiplets not involving the framing, the factor $-\zeta\left(u^{(a)}_j-u^{(a)}_i\right)=\zeta\left(u^{(a)}_i-u^{(a)}_j\right)$ from the vector multiplet would cancel the factor $\zeta\left(u^{(a)}_i-u^{(b)}_k-\epsilon_\chi\right)$ from the chiral multiplet when
\begin{equation}
    \label{u_ab}
    u^{(a)}_j=u^{(b)}_k+\epsilon_\chi\;.
\end{equation}
Likewise, the factor $-\zeta\left(u^{(c)}_j-u^{(a)}_i-\epsilon_\Lambda\right)=\zeta\left(u^{(a)}_i-u^{(c)}_j+\epsilon_\Lambda\right)$ from the Fermi multiplet would cancel the factor $\zeta\left(u^{(a)}_i-u^{(b)}_k-\epsilon_\chi\right)$ from the chiral multiplet when\footnote{Assume that $\mathfrak{b}_k\notin\partial_\chi\mathscr{M}$. It could be possible that there exist $\mathfrak{b}_k$ and $\mathfrak{b}'_{k'}$ such that $\epsilon(\mathfrak{b}_k)+\epsilon_\chi=\epsilon(\mathfrak{b}'_{k'})+\epsilon_{\chi'}=\epsilon(\mathfrak{c_j})-\epsilon_\Lambda$. Then $\zeta\left(u^{(a)}_i-u^{(c)}_j-\epsilon_\Lambda\right)$ would only cancel one such factor from the chiral multiplets. Nevertheless, there would also exist an atom $\mathfrak{a}_j$ with $\epsilon(\mathfrak{a}_j)=\epsilon(\mathfrak{b}_k)+\epsilon_\chi=\epsilon(\mathfrak{b}'_{k'})+\epsilon_{\chi'}$ so that the remaining factor would be cancelled by $\zeta\left(u^{(a)}_i-u^{(a)}_l\right)$ from the vector multiplet. If there are more than two such factors from the chiral multiplets for the same $\mathfrak{c}_j$ (and $\mathfrak{a}_l$), then there should exist other Fermi multiplets to cancel the extra factors due to the no-overlap condition.}
\begin{equation}
    \label{u_cb}
    u^{(c)}_j=u^{(b)}_k+\epsilon_\chi+\epsilon_\Lambda\;.
\end{equation}

It could sometimes be possible that there are two factors that $\zeta\left(u^{(a)}_i-u^{(a)}_j\right)$ could potentially cancel due to the $J$-/$E$-term relations\footnote{For non-toric cases, there could be more than two monomials of chiral multiplets in the corresponding $J$-terms (or $E$-terms), i.e., the Lagrangian contains a term $\Lambda(X_{i_1}\dots X_{i_l}+X_{j_1}\dots X_{j_m}+X_{k_1}\dots X_{k_n}+\dots)$. Nevertheless, due to the no-overlap condition, the extra factors should be cancelled by other factors from the Fermi multiplets as discussed here, so it suffices to consider the case with two such factors.}. Inside the crystal, such a situation can be depicted as follows:
\begin{equation}
    \includegraphics[width=3cm]{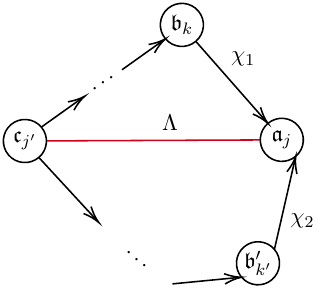}\;.\label{N=2cancellations}
\end{equation}
In this case, $\zeta\left(u^{(a)}_i-u^{(a)}_j\right)$ would only cancel one such factor, say, $\zeta\left(u^{(a)}_i-u^{(b)}_k-\epsilon_{\chi_1}\right)$. The other factor, $\zeta\left(u^{(a)}_i-u^{(b')}_{k'}-\epsilon_{\chi_2}\right)$, would be cancelled by the factor $\zeta\left(u^{(a)}_i-u^{(c)}_{j'}+\epsilon_\Lambda\right)$ from the Fermi multiplet.

Moreover, this also ensures that the contributions from the vector multiplet and the Fermi multiplet would not be used to cancel the same chiral multiplet in our procedure. This is because the $J$-/$E$-terms always involve at least two plaquettes as in \eqref{N=2cancellations}. As a result, if there is a vector multiplet and a Fermi multiplet giving the same factor (i.e., the factors from $\mathfrak{a}_j$ and $\mathfrak{c}_{j'}$ as above), then they would be used to cancel two different chirals (i.e., $\mathfrak{b}_k$ and $\mathfrak{b}'_{k'}$ as above). If there was only one such chiral (for example, $\mathfrak{b}_k\in\mathscr{M}$ while $\mathfrak{b}'_{k'}\notin\mathscr{M}$), then $\mathfrak{a}_j$ would not be in the molecule by the melting rule, and there would be no contribution $\zeta\left(u^{(a)}_i-u^{(a)}_j\right)$ from the vector multiplet doing the same cancellation as the Fermi multiplet\footnote{That $\mathfrak{a}_j\in\mathscr{M}$ and $\mathfrak{b}'_{k'}\notin\mathscr{M}$ could only be possible when $\mathfrak{a}_j$ is on certain boundary of the molecule (namely $\partial_\chi\mathscr{M}$ as defined in \eqref{boundaryN=2}). Then such factors from the vector multiplets would survive as analysed below.}.

We will find, similar to the $\mathcal{N}=4$ cases discussed above, that the non-cancelling contributions are associated with the boundaries of the crystal. The precise definition of the boundary, however, is different and is more involved for $\mathcal{N}=2$ cases. To begin with, for a given chiral multiplet $\chi$ and a Fermi multiplet $\Lambda$, and for two integer $m, n\in \mathbb{Z}$,
we define the associated boundary $\partial_{m \chi + n\Lambda } \mathscr{M}$ by
\begin{align}
\partial_{m \chi + n\Lambda } \mathscr{M}
:= \left\{
\mathfrak{a} \in \mathscr{M} \, | \, \epsilon(\mathfrak{a})
+ m\epsilon_\chi + n\epsilon_\Lambda \notin \mathscr{M}
\right\} \;.\label{boundaryN=2}
\end{align}

We are now ready to analyse situations where some of the factors inside the one-loop determinant would survive under the cancellations.

\paragraph{Uncancelled Chiral Multiplets}
    Let us consider a one-loop contribution from a chiral multiplet. For an atom $\mathfrak{b}_k$ connected by a chiral $\chi$, it is possible that there does not exist any atom $\mathfrak{a}_j$ or $\mathfrak{c}_j$ such that
    \begin{equation}
        \label{epsilon_abc}
        \epsilon(\mathfrak{a}_j)=\epsilon(\mathfrak{b}_k)+\epsilon_\chi\quad\text{or}\quad\epsilon(\mathfrak{c}_j)=\epsilon(\mathfrak{b}_k)+\epsilon_\chi+\epsilon_\Lambda\;.
    \end{equation}
    These equations mirror the situations in \eqref{u_ab} and \eqref{u_cb}, where $u_j^{(a)}$, $u_k^{(b)}$, $u^{(c)}_j$ are represented here by $\mathfrak{a}_j$, $\mathfrak{b}_k$, $\mathfrak{c}_j$, respectively.
    When these conditions are satisfied, we expect an uncancelled contribution from the chiral multiplet $\chi$.

    We now analyse when the two cases would happen.
        
    If $\mathfrak{b}_k\in\partial_\chi\mathscr{M}$, then this corresponds to the first case\footnote{If $\mathfrak{b}_k$ is both in $\partial_\chi\mathscr{M}$ and in $\partial_{\chi+\Lambda}\mathscr{M}$, then we can simply write $\mathfrak{b}_k\in\partial_\chi\mathscr{M}$ for the chiral factors that are not cancelled since $\partial_\chi\mathscr{M}\cap\partial_{\chi+\Lambda}\mathscr{M}\subseteq\partial_\chi\mathscr{M}$.}.
        This ensures that the chiral multiplet contribution is not cancelled by any vector multiplet contributions.

We still need to analyse the second case, i.e.\ cancellation against Fermi multiplets.
Suppose that we have $\mathfrak{b}_k\in\partial_\chi\mathscr{M}$.
If we denote the atom with position $\epsilon(\mathfrak{b}_k)+\epsilon_\chi$ as $\mathfrak{a}_l$, then we have $\mathfrak{a}_l\notin \mathscr{M}$ since $\mathfrak{b}_k\in\partial_\chi\mathscr{M}$. 
Let $\delta^c(\mathfrak{a}_l)$ (resp.~$\delta^F(\mathfrak{a}_l)$) be the number of atoms $\bullet$ in $\mathscr{M}$ such that $\epsilon(\bullet)+\epsilon_{\chi'}=\epsilon(\mathfrak{a}_l)$ (resp.~$\epsilon(\bullet)-\epsilon_{\Lambda'}=\epsilon(\mathfrak{a}_l)$) for some chiral $\chi'$ (resp.~Fermi $\Lambda'$).
These numbers count the number of chiral (and Fermi) multiplets ending on the atom $\mathfrak{a}_l$, each of which generates a factor in the denominator (numerator).

\begin{figure}[ht]
    \centering
    \includegraphics[width=5cm]{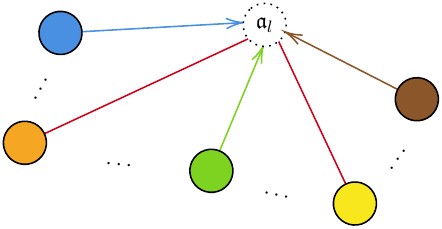}
    \caption{In the figure, suppose that all the atoms are in the molecule, except $\mathfrak{a}_l$ as indicated by the dashed circle. We have $\epsilon(\mathfrak{a}_l)=\epsilon({\color{blue} \bullet})+{\color{blue} \chi}=\epsilon({\color{orange} \bullet})-{\color{orange} \Lambda}=\epsilon({\color{green} \bullet})+{\color{green} \chi}=\epsilon({\color{yellow} \bullet})-{\color{yellow} \Lambda}=\epsilon({\color{brown} \bullet})+{\color{brown} \chi}$. Then there is a factor $\zeta(\mathfrak{a}_i-{\color{blue} \bullet}-\epsilon_{\color{blue} \chi})^{1/3}\zeta(\mathfrak{a}_i-{\color{green} \bullet}-\epsilon_{\color{green} \chi})^{1/3}\zeta(\mathfrak{a}_i-{\color{brown} \bullet}-\epsilon_{\color{brown} \chi})^{1/3}$ in the denominator with $\delta_{\color{blue} \chi}({\color{blue} \bullet})=\delta_{\color{green} \chi}({\color{green} \bullet})=\delta_{\color{brown} \chi}({\color{brown} \bullet})=1/3$ (for brevity, we have used the atom to denote its coordinates in $\zeta$). If the brown atom is not in the molecule, then there are no such factors as $\delta^c(\mathfrak{a}_l)=\delta^F(\mathfrak{a}_l)$. If the brown and the yellow atoms are not in the molecule (which is possible if some atom (with its precedent chiral $\chi'$) between the green and the yellow atoms is in the boundary $\partial_{-\chi'}\mathscr{M}$ of the molecule), then there is a factor $\zeta(\mathfrak{a}_i-{\color{blue} \bullet}-\epsilon_{\color{blue} \chi})^{1/2}\zeta(\mathfrak{a}_i-{\color{green} \bullet}-\epsilon_{\color{green} \chi})^{1/2}$ in the denominator with $\delta_{\color{blue} \chi}({\color{blue} \bullet})=\delta_{\color{green} \chi}({\color{green} \bullet})=1/2$. Notice that all such factors are equal to $\zeta(\mathfrak{a}_j-\mathfrak{a}_l)$, and hence always give integer powers in total. It could also be possible that, for example, the brown and the green atoms are not in the molecule. Then $\delta_{\color{blue} \chi}({\color{blue} \bullet})=0$ due to the contribution from ${\color{orange} \Lambda}$, and the factor from ${\color{yellow} \Lambda}$ would survive as it belongs to $\partial_{-\cup\chi-\Lambda}\mathscr{M}$ defined below in \eqref{capchiLambda}.}\label{deltaillustration}
\end{figure}

Since we always assume the no-overlap condition (and the generic framing), there would be no poles of order greater than one, and $\delta^c(\mathfrak{a}_l)$ can never be greater than $\delta^F(\mathfrak{a}_l)+1$. This means that, as explained and illustrated in \cref{deltaillustration}, there would be one (resp.~no) surviving chiral factor when $\delta^c(\mathfrak{a}_l)=\delta^F(\mathfrak{a}_l)+1$ (resp.~$\delta^c(\mathfrak{a}_l)\leq\delta^F(\mathfrak{a}_l)$)\footnote{For non-generic framing, the no-overlap condition may also be satisfied with general $\delta^c(\mathfrak{a}_l)>\delta^F(\mathfrak{a}_l)$ as the higher order poles may be cancelled by some framing arrows. This corresponds to the pausers and the stoppers in \cite{Galakhov:2021xum}. Then $\delta_\chi(\mathfrak{a}_l)$ should be modified to $\left(\delta^c(\mathfrak{a}_l)-\delta^F(\mathfrak{a}_l)\right)/\delta^c(\mathfrak{a}_l)$. We shall always consider the generic framing in what follows.}. 
 If there is one such surviving chiral factor, one obvious choice is to pick up one particular chiral as representing the uncanceled factor. This, however, requires an arbitrary choice of such a chiral multiplet. We can instead choose to treat all the chiral on equal footing
and write the remaining factor as the product of all the $\delta^c(\mathfrak{a}_l)$ chiral factors but with the power $1/\delta^c(\mathfrak{a}_l)$---schematically, we have $\zeta=(\zeta\zeta\dots\zeta)^{1/\delta^c(\mathfrak{a}_l)}$. This motivates us to define a number
$\delta_\chi(\mathfrak{b}_k)$ by
\begin{align}
    \delta_\chi(\mathfrak{b}_k) 
    := \begin{cases}
        1/\delta^c(\mathfrak{a}_l) & \delta^c(\mathfrak{a}_l)>\delta^F(\mathfrak{a}_l) \\
        0 & \delta^c(\mathfrak{a}_l)\leq\delta^F(\mathfrak{a}_l) \;,
    \end{cases}
\end{align}
and we will then have an uncancelled factor for each chiral multiplet $\chi$, 
with power given by $\delta_\chi(\mathfrak{b}_k)$. Note that 
if there are no surviving chiral factors, then $\delta(\mathfrak{b}_j)=0$ gives $\zeta^0=1$, as it should. Collecting these factors, one obtains the total uncancelled expression to be
\begin{align}
    &\mathcal{Z}_\chi(\bm{\epsilon})=\prod_{a,b\in Q_0}\prod_{\chi\in\{b\rightarrow a\}}\prod_{\substack{\mathfrak{a}_i\in\mathscr{M}\\\mathfrak{b}_j\in\partial_\chi\mathscr{M}}}\frac{1}{E_{ij}(-\epsilon_\chi)^{\delta_\chi(\mathfrak{b}_j)}}\;,
\end{align}

\paragraph{Uncancelled Vector Multiplets}

Let us next consider a vector multiplet one-loop determinant associated with an atom $\mathfrak{a}_j$.
For the atom $\mathfrak{a}_j$, it is possible that there does not exist any atom $\mathfrak{b}_k$ such that
    \begin{equation}
        \epsilon(\mathfrak{b}_k)=\epsilon(\mathfrak{a}_j)-\epsilon_\chi,
    \end{equation}
    which mirrors the situation in \eqref{u_ab}, where $u_j^{(a)}, u_k^{(b)}$ are represented here by $\mathfrak{a}_j, \mathfrak{b}_k$.
 we then expect an uncancelled contribution from a vector multiplet.
This happens when $\mathfrak{a}_j\in\partial_{-\chi}\mathscr{M}$, and hence we have
the factor
\begin{align}
    &\mathcal{Z}_V(\bm{\epsilon})=\prod_{a,b\in Q_0}\prod_{\substack{\mathfrak{a}_i\in\mathscr{M}\\\mathfrak
    {a}_i\neq\mathfrak{a}_j\in\bigcap\limits_{\chi\in\{b\rightarrow a\}}\partial_{-\chi}\mathscr{M}}}E_{ij}(0)\;,
\end{align}
Notice that some factors with $\mathfrak{a}_{j}$ being initial atoms would be cancelled by the factors from the framing. However, to keep our expression simpler (as not all $a\in Q_0$ would correspond to initial atoms and different sub-molecules may have different $a$), we have kept them in the formula above.

\paragraph{Uncancelled Fermi Multiplets}

Let us next discuss a Fermi multiplet ending on an atom $\mathfrak{c}_j$ in the crystal.
For the atom $\mathfrak{c}_j$, it is possible that there does not exist any $\mathfrak{b}_k$ such that
    \begin{equation}
        \quad\epsilon(\mathfrak{b}_k)=\epsilon(\mathfrak{c}_j)-\epsilon_\chi-\epsilon_\Lambda\;.
    \end{equation}
    Again, these equations mirror the situation in \eqref{u_cb}, where $u_k^{(b)}, u^{(c)}_j$ are represented here by $\mathfrak{b}_k, \mathfrak{c}_j$, respectively.
    When this condition is satisfied, we expect an uncancelled contribution from the Fermi multiplet $\Lambda$.
    This happens when $\mathfrak{c}_j\in\partial_{-\chi-\Lambda}\mathscr{M}$. By collecting uncancelled contributions, one obtains
\begin{align}
   &\mathcal{Z}_\Lambda(\bm{\epsilon})=\prod_{a,b\in Q_0}\prod_{\Lambda\in\{b\rightarrow a\}}\prod_{\substack{\mathfrak{a}_i\in\mathscr{M}\\\mathfrak{b}_j\in\partial_{-\cup\chi-\Lambda}\mathscr{M}}}E_{ij}(\epsilon_\Lambda)\;,
\end{align}
where we have defined 
\begin{align}
\partial_{-\cup\chi-\Lambda}\mathscr{M}:=
\bigcap\limits_{\substack{\chi\in I_c\\t(\chi)=t(\Lambda)}}\partial_{-\chi-\Lambda}\mathscr{M}\label{capchiLambda}
\end{align}
to account for contributions from different chiral multiplets $\chi$.

\subsection{The \texorpdfstring{$\mathcal{N}=2$}{N=2} Formula}\label{subsec:N2_formula}

We are now ready to write down our main formula for $\mathcal{N}=2$ grand-canonical partition function, by collecting the factors described above:
\begin{tcolorbox}[ams align]
    \mathcal{Z}(\bm{q}, \bm{\epsilon})=\sum_{\mathscr{M}\subset \mathscr{C}}\bm{q}^{\bm{N}}\xi \, \mathcal{Z}_V(\bm{\epsilon})\, \mathcal{Z}_\chi(\bm{\epsilon})\, \mathcal{Z}_\Lambda(\bm{\epsilon})\, \mathcal{Z}_\mathfrak{f}(\bm{\epsilon})\;,
\end{tcolorbox}
\noindent
with
\begin{subequations}\label{ZsN=2}
\begin{tcolorbox}[ams align]
    &\mathcal{Z}_V(\bm{\epsilon})=\prod_{a,b\in Q_0}\prod_{\substack{\mathfrak{a}_i\in\mathscr{M}\\\mathfrak
    {a}_i\neq\mathfrak{a}_j\in\bigcap\limits_{\chi\in\{b\rightarrow a\}}\partial_{-\chi}\mathscr{M}}}E_{ij}(0)\;,\\
    &\mathcal{Z}_\chi(\bm{\epsilon})=\prod_{a,b\in Q_0}\prod_{\chi\in\{b\rightarrow a\}}\prod_{\substack{\mathfrak{a}_i\in\mathscr{M}\\\mathfrak{b}_j\in\partial_\chi\mathscr{M}}}\frac{1}{E_{ij}(-\epsilon_\chi)^{\delta_\chi(\mathfrak{b}_j)}}\;,\\
    &\mathcal{Z}_\Lambda(\bm{\epsilon})=\prod_{a,b\in Q_0}\prod_{\Lambda\in\{b\rightarrow a\}}\prod_{\substack{\mathfrak{a}_i\in\mathscr{M}\\\mathfrak{b}_j\in\partial_{-\cup\chi-\Lambda}\mathscr{M}}}E_{ij}(\epsilon_\Lambda)\;,\\
    &\mathcal{Z}_\mathfrak{f}(\bm{\epsilon})=\prod_{a\in Q_0}\left(\left(\frac{\prod\limits_{\Lambda\in\{\infty\rightarrow a\}}\prod\limits_{\mathfrak{a}_i\in\mathscr{M}}E_{00}(v_\Lambda-\epsilon(\mathfrak{a}_i))}{\prod\limits_{\chi\in\{\infty\rightarrow a\}}\prod\limits_{\mathfrak{a}_i\in\mathscr{M}\backslash\mathscr{O}}E_{00}(\epsilon(\mathfrak{a}_i)-v_\chi)}\right)\left(\frac{\prod\limits_{\Lambda\in\{a\rightarrow\infty\}}\prod\limits_{\mathfrak{a}_i\in\mathscr{M}}E_{00}(\epsilon(\mathfrak{a}_i)-v_\Lambda)}{\prod\limits_{\chi\in\{a\rightarrow\infty\}}\prod\limits_{\mathfrak{a}_i\in\mathscr{M}}E_{00}(v_\chi-\epsilon(\mathfrak{a}_i))}\right)\right)\;.
\end{tcolorbox}
\end{subequations}

We can avoid the fractional powers $\delta_\chi(\mathfrak{b}_j)$ by rewriting the partition function. We define the non-trivial outer boundary $\overline{\partial}_a\mathscr{M}$ for $a \in Q_0$ as
\begin{equation}
    \overline{\partial}_a\mathscr{M}:=\left\{\mathfrak{a}\in\overline{\partial}_\text{full}\mathscr{M}\;|\;\delta^c(\mathfrak{a})=\delta^F(\mathfrak{a})+1\right\}\;,
\end{equation}
where $\delta^c(\mathfrak{a})$, $\delta^F(\mathfrak{a})$ were defined above as in \cref{subsec:cancellation_N2}, and $\overline{\partial}_\text{full}\mathscr{M}=\{\mathfrak{a}\notin\mathscr{M}\;|\;\epsilon(\mathfrak{a})=\epsilon(\mathfrak{b})+\epsilon_\chi,\;\exists\;b\in Q_0,~\mathfrak{b}\in\mathscr{M},\text{ and }\chi\in I_c\}$ is the full outer boundary of $\mathscr{M}$. Then $\mathcal{Z}_\chi$ may be rewritten as
\begin{equation}
    \mathcal{Z}_\chi(\bm{\epsilon})=\prod_{a\in Q_0}\prod_{\substack{\mathfrak{a}_i\in\mathscr{M}\\\mathfrak{a}_j\in\overline{\partial}_a\mathscr{M}}}\frac{1}{E_{ij}(0)}\;.
\end{equation}
Likewise, we may define another outer boundary $\overline{\partial}_{-a}\mathscr{M}$ ($a\in Q_0$) to be the set
\begin{equation}
\overline{\partial}_{-a}\mathscr{M}:=
    \bigcup_{b\in Q_0}\left(\left(\bigcap_{\chi\in\{b\rightarrow a\}}\partial_{-\chi}\mathscr{M}\right)\bigcup\left(\bigcup_{\Lambda\in\{b\rightarrow a\}}\{\mathfrak{a}\notin\mathscr{M}\;|\;\epsilon(\mathfrak{a})=\epsilon(\mathfrak{b})-\epsilon_\Lambda,\;\mathfrak{b}\in\partial_{-\cup\chi-\Lambda}\mathscr{M}\}\right)\right)\;.
\end{equation}
Then the partition function can be written as
\begin{subequations}
\begin{tcolorbox}[ams align]
    &\mathcal{Z}(\bm{q}, \bm{\epsilon})=\sum_{\mathscr{M}\subset \mathscr{C}}\bm{q}^{\bm{N}}\xi \, \mathcal{Z}_{\overline{\partial}}(\bm{\epsilon})\, \mathcal{Z}_\mathfrak{f}(\bm{\epsilon})\;,\\
    &\mathcal{Z}_{\overline{\partial}}(\bm{\epsilon})=\prod_{a\in Q_0}\frac{\prod\limits_{\substack{\mathfrak{a}_i\in\mathscr{M}\\\mathfrak{a}_i\neq\mathfrak{a}_j\in\overline{\partial}_{-a}\mathscr{M}}}E_{ij}(0)}{\prod\limits_{\substack{\mathfrak{a}_i\in\mathscr{M}\\\mathfrak{a}_j\in\overline{\partial}_a\mathscr{M}}}E_{ij}(0)}\;,\\
    &\mathcal{Z}_\mathfrak{f}(\bm{\epsilon})=\prod_{a\in Q_0}\left(\left(\frac{\prod\limits_{\Lambda\in\{\infty\rightarrow a\}}\prod\limits_{\mathfrak{a}_i\in\mathscr{M}}E_{00}(v_\Lambda-\epsilon(\mathfrak{a}_i))}{\prod\limits_{\chi\in\{\infty\rightarrow a\}}\prod\limits_{\mathfrak{a}_i\in\mathscr{M}\backslash\mathscr{O}}E_{00}(\epsilon(\mathfrak{a}_i)-v_\chi)}\right)\left(\frac{\prod\limits_{\Lambda\in\{a\rightarrow\infty\}}\prod\limits_{\mathfrak{a}_i\in\mathscr{M}}E_{00}(\epsilon(\mathfrak{a}_i)-v_\Lambda)}{\prod\limits_{\chi\in\{a\rightarrow\infty\}}\prod\limits_{\mathfrak{a}_i\in\mathscr{M}}E_{00}(v_\chi-\epsilon(\mathfrak{a}_i))}\right)\right)\;.
\end{tcolorbox}
\end{subequations}

\subsection{Comparison between
\texorpdfstring{$\mathcal{N}=4$}{N=4} and \texorpdfstring{$\mathcal{N}=2$}{N=2} Formulas}

We now comment on the comparison between $\mathcal{N}=4$ and $\mathcal{N}=2$ formulas.
While the overall structure for the $\mathcal{N}=2$ formula is similar to the $\mathcal{N}=4$ formula, 
there are important differences between the two. On the one hand, the expressions are simpler since the one-loop determinants are themselves simpler; this is because $\mathcal{N}=2$ multiplets are smaller than their $\mathcal{N}=4$ counterparts. On the other hand, 
the specification of uncancelled contributions, and relatedly of the boundary of the molecule, is more involved.

Since an $\mathcal{N}=4$ theory is automatically an $\mathcal{N}=2$ theory, we have two different-looking expressions even if we set the extra fugacity $\varepsilon=0$. Of course, these two formulas should coincide since we are simply evaluating the same JK residue with two different cancellation schemes.

In an $\mathcal{N}=4$ theory, the $\mathcal{N}=2$ Fermis and the chirals are paired to form $\mathcal{N}=4$ chirals (and likewise for the vector multiplets). Then the weight of Fermi can be written as $\epsilon_\Lambda=-\epsilon+\epsilon_\chi$ with $\chi$ being its pair. Recall that the factor $\zeta\left(u^{(a)}_i-u^{(b)}_j-\epsilon_\chi\right)$ in the denominator is cancelled by $-\zeta\left(u^{(a)}-u^{(b)}_k+\epsilon-\epsilon_\chi\right)$ in the $\mathcal{N}=4$ cases in \cref{subsec:cancellation_N4}, where $u^{(b)}_k=u^{(b)}_j+\epsilon$. In the $\mathcal{N}=2$ cases in \cref{subsec:cancellation_N2}, the chiral $\chi$ with $\zeta\left(u^{(a)}_i-u^{(b)}_j-\epsilon_\chi\right)$ would in contrast be cancelled by the Fermi $\Lambda'$ with $-\zeta\left(u^{(c)}_k-u^{(a)}_i+\epsilon-\epsilon_{\chi'}\right)=\zeta\left(u^{(a)}_i-u^{(c)}_k-\epsilon+\epsilon_{\chi'}\right)$, where $u^{(c)}_k=u^{(b)}_j+\epsilon_\chi-\epsilon+\epsilon_{\chi'}$ (and we have used $\epsilon_{\Lambda'}=-\epsilon+\epsilon_{\chi'}$)\footnote{Moreover, when decomposing the $\mathcal{N}=4$ cases to the $\mathcal{N}=2$ language, the framing would often be non-generic. This is in fact an important feature since it would give rise to pausers and/or stoppers so that the $\mathcal{N}=4$ and the $\mathcal{N}=2$ expressions would lead to the same crystal structure as expected.}. The two different cancellation mechanisms indicate that we have a non-trivial reshuffling of factors between the $\mathcal{N}=4$ formula and the $\mathcal{N}=2$ formula (with the non-genericity of the framing taken into account).

We note in passing that the concept of boundaries is different for $\mathcal{N}=2$ and $\mathcal{N}=4$ theories. 
here, the counterpart of 
$\partial_+ \mathscr{M}$ (resp.~$\partial_- \mathscr{M}$) is played by a collection of boundaries
$\partial_{\chi} \mathscr{M}$
and $\partial_{-\chi-\Lambda} \mathscr{M}$ (resp.~$\partial_{-\chi} \mathscr{M}$
and $\partial_{\chi+\Lambda} \mathscr{M}$).
It is important to note, however, that the relevant boundaries are different between $\mathcal{N}=4$ and $\mathcal{N}=2$.
For example, it is easy to come up with examples where 
\begin{equation}
    \bigcup_{\chi\in I_c}\partial_\chi\mathscr{M}\subsetneq\partial_+\mathscr{M}\;,
\end{equation}
and likewise for $\partial_{-\chi}\mathscr{M}$ and $\partial_-\mathscr{M}$. Suppose that a molecule is given by the Young diagram (where the two directions of the Young diagram are given by the two chiral multiplets $\chi_1$ and $\chi_2$ respectively) with three atoms of coordinates $v_1$, $v_1+\epsilon_{\chi_1}$, $v_1+\epsilon_{\chi_2}$. All three atoms belong to $\partial_+\mathscr{M}$, but the one with $v_1$ does not belong to $\partial_{\chi_1}\mathscr{M}$ or $\partial_{\chi_2}\mathscr{M}$, assuming $\varepsilon=\epsilon_1 +\epsilon_2 +\dots$.

\section{Examples}\label{examples}
In this section, let us discuss some examples. We shall consider one $\mathcal{N}=4$ example and two $\mathcal{N}=2$ examples.

In our discussion, we will encounter expressions
that are reminiscent of those from the combinatorics of Young diagrams. To make this connection more explicit, let us 
define the $\chi$-leg length $L_{\chi,\mathscr{M}_n}(\mathfrak{a})$ for an atom $\mathfrak{a}$ in the $m^\text{th}$ sub-molecule $\mathscr{M}_m$ to be
the ``distance from $\mathfrak{a}$ to the boundary of $\mathscr{M}_m$ in the direction of $\chi$''. In equations, we define
\begin{equation}
    L_{\chi,\mathscr{M}_n}(\mathfrak{a})
    :=\frac{1}{\epsilon_\chi}\left(\epsilon(\mathfrak{b})-\epsilon(\mathfrak{a})\right)\;,
\end{equation}
where we write $\epsilon(\mathfrak{a})=v_{l_m}+\sum\limits_{x\in I_c} k_x\epsilon_x$, and the atom $\mathfrak{b}$
is the atom on the boundary of $\mathscr{M}_m$, satisfying
\begin{equation}
    \epsilon(\mathfrak{b})=v_{l_n}+k'_\chi\epsilon_\chi+\sum_{x\neq\chi}k_x\epsilon_x\in\partial_\chi\mathscr{M}_n\quad (k'_{\chi}\ge k_{\chi})\;.
\end{equation}

\subsection{An \texorpdfstring{$\mathcal{N}=4$}{N=4} Example}\label{N=4ex}
As an $\mathcal{N}=4$ example, let us take the Jordan quiver:
\begin{equation}
    \includegraphics[width=3.5cm]{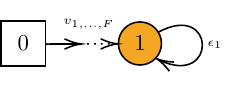}.
\end{equation}
The one-loop determinant reads
\begin{equation}
    Z_\text{1-loop}=Z_VZ_\text{adj}Z_\mathfrak{f}\;,
\end{equation}
where\footnote{When $i=j$, the factors in $Z_V$ are understood as $-1/\zeta(\epsilon)$.}
\begin{equation}
    Z_V=[\text{d}\bm{u}] \, \xi(N)\prod_{i,j}\frac{-\zeta(u_i-u_j)}{\zeta(u_i-u_j+\varepsilon)}\;,\quad Z_\text{adj}=\prod_{i,j}\frac{-\zeta(u_i-u_j+\varepsilon-\epsilon_1)}{\zeta(u_i-u_j-\epsilon_1)}\;,\quad Z_\mathfrak{f}=\prod_{n=1}^F\prod_{i=1}^N\frac{-\zeta(u_i-v_n+\varepsilon)}{\zeta(u_i-v_n)}\;.\label{ZsJordan}
\end{equation}
The molecule would be $F$ disjoint one-dimensional arrays of atoms, with a total number of $N$ atoms \cite{Bao:2025hfu}:
\begin{equation}
    \includegraphics[width=5cm]{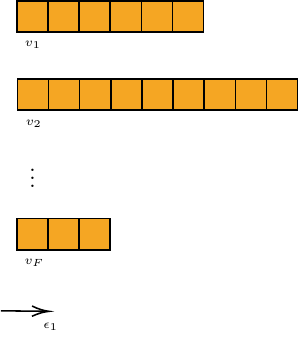}\;.
\end{equation}
The direction $\epsilon_{\chi_1}$ may then be written as $\epsilon_1$.

The grand-canonical partition function reads
\begin{equation}
    \mathcal{Z}(q)
    =\sum_{\mathscr{M}}q^N \xi\, \mathcal{Z}_V\, \mathcal{Z}_\chi\, \mathcal{Z}_\mathfrak{f}\;,
\end{equation}
where\footnote{In this case, we need to keep $\epsilon\neq\epsilon_1$ in the refinement as there is only one parameter $\epsilon_1$. Therefore, all the atoms are in $\partial_+\mathscr{M}$ and $\partial_-\mathscr{M}$.}
\begin{subequations}
\begin{align}
    \mathcal{Z}_V&=\prod_{\mathfrak{a}_i\in\mathscr{M}}\frac{\prod\limits_{\substack{\mathfrak{b}_j\in\partial_+\mathscr{M}\\j\neq i}}E_{ij}(0)}{\prod\limits_{\substack{\mathfrak{b}_j\in\partial_-\mathscr{M}\\j\neq i}}E_{ij}(\varepsilon)}\;,\\
    \mathcal{Z}_\chi&=(-1)^N\prod_{\mathfrak{a}_i\in\mathscr{M}}\frac{\prod\limits_{\mathfrak{b}_j\in\partial_-\mathscr{M}}E_{ij}(\varepsilon-\epsilon_1)}{\prod\limits_{\substack{\mathfrak{b}_j\in\partial_+\mathscr{M}\\\epsilon(\mathfrak{a}_i)+\epsilon_1\neq\epsilon(\mathfrak{a}_j)}}E_{ij}(-\epsilon_1)}\;,\\
    \mathcal{Z}_\mathfrak{f}&=(-1)^{FN}\prod_{n=1}^F\frac{\prod\limits_{\mathfrak{a}_i\in\partial_+\mathscr{M}}E_{00}(\epsilon(\mathfrak{a}_i)-v_n+\varepsilon)}{\prod\limits_{\mathfrak{a}_i\partial_-\mathscr{M}\backslash\mathscr{O}}E_{00}(\epsilon(\mathfrak{a}_i)-v_n)}\;.
\end{align}
\end{subequations}
Given an atom $\mathfrak{a}$ of position $k_m\epsilon_1$ in the sub-molecule (notice that $k_m$ starts from 0), recall that its leg length $L_{\mathscr{M}_n}(\mathfrak{a})$ in the sub-molecule $\mathscr{M}_n$ is given by
\begin{equation}
    L_{\mathscr{M}_n}(\mathfrak{a})=N_n-k_m\;,
\end{equation}
where $N_n$ is the number of atoms in $\mathscr{M}_n$. The partition function can be further simplified to
\begin{equation}
    \mathcal{Z}=\sum_\mathscr{M}q^N\xi(N) (-1)^{FN}\prod_{m,n=1}^F\prod_{\mathfrak{a}\in\mathscr{M}_m}\frac{\zeta(\varepsilon-L_{\mathscr{M}_n}(\mathfrak{a})\epsilon_1+v_m-v_n)}{\zeta(L_{\mathscr{M}_n}(\mathfrak{a})\epsilon_1+v_m-v_n)}\;.
\end{equation}
This is the well-known vortex partition function with an adjoint matter. The unrefined limit can be obtained by first taking $\varepsilon\rightarrow0$ and then taking $\epsilon_1\rightarrow0$. The unrefined partition function is
\begin{equation}
    \mathcal{Z}=\sum_{\mathscr{M}}(-1)^{FN}\xi(N)q^N\;.
\end{equation}

Let us also try to use the $\mathcal{N}=2$ language to obtain the partition function. The two $\mathcal{N}=2$ chirals (with weights $\epsilon_1$ and $-\epsilon$) seem to give a 2-dimensional crystal instead. However, as mentioned above, the framing would be non-generic. The Fermis from the framing would cancel the factors from the chiral with weight $-\epsilon$, which provides stoppers to the crystals and reduces it back to the above 1-dimensional structure\footnote{The $\mathcal{N}=2$ chiral comes from the $\mathcal{N}=4$ vector multiplet, and this is consistent with the fact that the poles from the vector multiplet do not contribute in the JK residue formula in this case.}. Although the framing is non-generic, we may still use the $\mathcal{N}=2$ cancellation mechanism. It turns out that the denominator (resp.~numerator) only has the contribution from the addable atoms (resp.~initial atoms). Writing the positions of these atoms in terms of the leg lengths, the partition function reads
\begin{equation}
    \mathcal{Z}=\sum_\mathscr{M}q^N\xi(N)\prod_{m,n=1}^F\prod_{\mathfrak{a}\in\mathscr{M}_m}\frac{-\zeta(\varepsilon-L_{\mathscr{M}_n}(\mathfrak{a})\epsilon_1+v_m-v_n)}{\zeta(L_{\mathscr{M}_n}(\mathfrak{a})\epsilon_1+v_m-v_n)}\;.
\end{equation}
In the numerator, there are $N_1F+N_2F+\dots+N_FF$ minus signs (where $N_i$ denotes the number of atoms in the $n^\text{th}$ sub-molecule). This gives an overall sign $(-1)^{FN}$, and we have the partition function as expected.

\subsection{An \texorpdfstring{$\mathcal{N}=2$}{N=2} Example}\label{N=2ex}
Let us now consider an $\mathcal{N}=2$ example:
\begin{equation}
    \includegraphics[width=5cm]{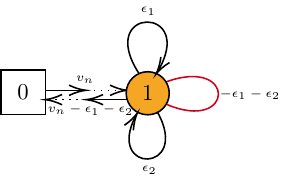}.
\end{equation}
There are $F$ chiral multiplets from the framing node to the gauge node of weights $v_n$ ($n=1,\dots, F$). Moreover, there are $F$ chiral multiplets from the gauge node to the framing node of weights $v_n-\epsilon_1-\epsilon_2$. The one-loop determinant is
\begin{equation}
    Z_\text{1-loop}=Z_V\, Z_\chi \, Z_\Lambda \, Z_\mathfrak{f}\;,
\end{equation}
where
\begin{align}
    &Z_V=[\text{d}\bm{u}] \, \xi(N)\prod_{i\neq j}^N(-\zeta(u_i-u_j))\;,\quad Z_\chi=\prod_{\alpha=1}^2\prod_{i,j}^N\frac{1}{\zeta(u_i-u_j-\epsilon_\alpha)}\;,\nonumber\\
    &Z_\Lambda=\prod_{i,j}^N(-\zeta(u_i-u_j+\epsilon_1+\epsilon_2))\;,\quad Z_\mathfrak{f}=\prod_{i=1}^N\frac{1}{\zeta(u_i-v_1)\zeta(v_1-\epsilon_1-\epsilon_2-u_i)}\;.
\end{align}
The molecules are in fact given by the 2d Young diagrams with a total number of $N$ atoms \cite{Nekrasov:2002qd,Losev:2003py,Nakajima:2003pg,Nekrasov:2003rj,Bao:2025hfu,Jiang:2025luj}:
\begin{equation}
    \includegraphics[width=10cm]{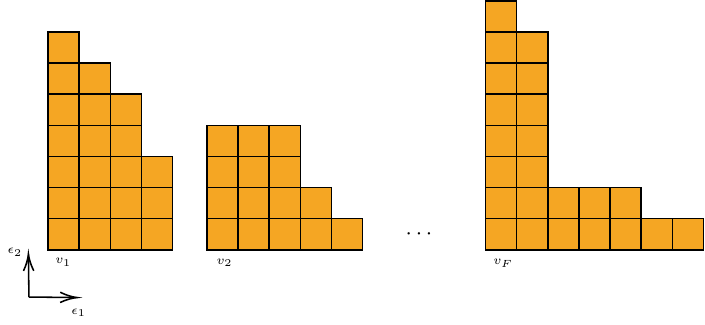}\;.
\end{equation}
Therefore, in this case, there is a standard choice of the directions and the basis $\{\epsilon_\alpha\}$ with $\epsilon_1$ and $\epsilon_2$ corresponding to the two arrows $\epsilon_{\chi_1}$ and $\epsilon_{\chi_2}$ respectively. We may write $\partial_{\chi_\alpha}\mathscr{M}$ as $\partial_\alpha\mathscr{M}$ for $\alpha=1,2$. The Fermi multiplet corresponds to the diagonal direction $-\epsilon_1-\epsilon_2$.

The partition function reads
\begin{equation}
    \mathcal{Z}=\sum_{\mathscr{M}}\bm{q}^N \xi \, \mathcal{Z}_V\, \mathcal{Z}_\chi \, \mathcal{Z}_\Lambda \, \mathcal{Z}_\mathfrak{f}\;,
\end{equation}
where
\begin{subequations}
\begin{align}
    \mathcal{Z}_V=&\prod_{n=1}^F\prod_{\mathfrak{a}_i\in\mathscr{M}\backslash\mathscr{O}}E_{00}(\epsilon(\mathfrak{a}_i)-v_n)\;,\\
    \mathcal{Z}_\chi=&\prod_{\alpha=1}^2\prod_{\substack{\mathfrak{a}_i\in\mathscr{M}\\\mathfrak{a}_j\in\partial\mathscr{M}}}\frac{1}{E_{ij}(-\epsilon_\alpha)^{\delta_\alpha(\mathfrak{a}_j)}}\;,\\
    \mathcal{Z}_\Lambda=&\prod_{\mathfrak{a}_i\in\mathscr{M}}\prod_{\substack{\mathfrak{a}_i\in\mathscr{M}\\\mathfrak{a}_j\in\text{Rem}(\mathscr{M})}}E_{ij}(-\epsilon_1-\epsilon_2)\;,\\
    \mathcal{Z}_\mathfrak{f}=&\prod_{n=1}^F\left(\left(\prod_{\mathfrak{a}_i\in\mathscr{M}\backslash\mathscr{O}}\frac{1}{E_{00}(\epsilon(\mathfrak{a}_i)-v_n)}\right)\left(\prod_{\mathfrak{a}_i\in\mathscr{M}}\frac{1}{E_{00}(v_n-\epsilon_1-\epsilon_2-\epsilon(\mathfrak{a}_i))}\right)\right)\;.
\end{align}
\end{subequations}
Here, $\text{Rem}(\mathscr{M})$ is the set of removable atoms of $\mathscr{M}$ (while preserving the condition that the molecule is described by the Young diagram). This is because the set of atoms in $\partial_{-\cup\chi-\Lambda}$ satisfies $\epsilon(\mathfrak{a})+\epsilon_1\notin\mathscr{M}$ and $\epsilon(\mathfrak{a})+\epsilon_2\notin\mathscr{M}$. In other words, this is the set of atoms at the outer corner of the profile of the Young diagrams, which are precisely the removable atoms.

It is clear that the factors in $\mathcal{Z}_V$ get cancelled by the corresponding factors in $\mathcal{Z}_\mathfrak{f}$. Moreover, $\delta_\alpha(\mathfrak{a}_j)=1/2$ if $\epsilon(\mathfrak{a}_j)+\epsilon_\alpha\notin\mathscr{M}$ and $\epsilon(\mathfrak{a}_j)-\epsilon_\beta$ is at the inner corner of the profile of the Young diagrams ($\beta\neq\alpha$). If $\mathfrak{a}_j$ is on the shadow of the Young diagrams, then $\delta_\alpha(\mathfrak{a}_j)=1$. Otherwise, $\delta_\alpha(\mathfrak{a}_j)=0$. In other words, $\mathcal{Z}_\chi$ can be written as
\begin{equation}
    \mathcal{Z}_\chi=\prod_{\substack{\mathfrak{a}_i\in\mathscr{M}\\\mathfrak{a}_j\in\text{Add}(\mathscr{M})}}\frac{1}{E_{ij}(0)}\;,
\end{equation}
where $\text{Add}(\mathscr{M})$ is the set of addable atoms of $\mathscr{M}$.

This is exactly the shell formula for the Young diagrams in \cite{Jiang:2025luj}. As shown in \cite[Appendix C.1]{Jiang:2025luj}, this expression is equal to the formula in \cite{Nekrasov:2002qd,Losev:2003py,Nakajima:2003pg,Nekrasov:2003rj} using the arm and the leg lengths:
\begin{align}
    \mathcal{Z}=\sum_{\mathscr{M}}q^N \xi(N)\prod_{m,n=1}^F&\prod_{\mathfrak{a}\in\mathscr{M}_m}\frac{1}{\zeta((1+L_{\mathscr{M}_m}(\mathfrak{a}))\epsilon_1-A_{\mathscr{M}_m}(\mathfrak{a})\epsilon_2+v_m-v_n)}\nonumber\\
    &\prod_{\mathfrak{a}\in\mathscr{M}_n}\frac{1}{\zeta(-L_{\mathscr{M}_n}(\mathfrak{a})\epsilon_1+(1+A_{\mathscr{M}_n}(\mathfrak{a}))\epsilon_2+v_m-v_n)}\;.
\end{align}
Here, the leg (resp.~arm) refers to the horizontal (resp.~vertical) direction, that is, $L_\mathscr{M}(\mathfrak{a})=L_{1,\mathscr{M}}(\mathfrak{a})$ and $A_\mathscr{M}(\mathfrak{a})=L_{2,\mathscr{M}}(\mathfrak{a})$.

\subsection{Another \texorpdfstring{$\mathcal{N}=2$}{N=2} Example: \texorpdfstring{$Q^{1,1,1}$}{Q111}}\label{Q111}
Let us consider one of the toric phases of the toric geometry known as $Q^{1,1,1}$ \cite{Franco:2015tna,Franco:2015tya}. The quiver is
\begin{equation}
    \includegraphics[width=5cm]{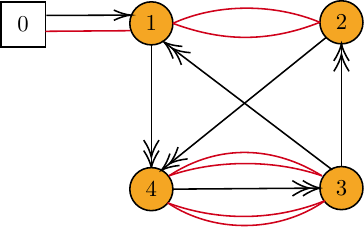}\;.
\end{equation}
The weights of the edges are
\begin{equation}
    \begin{tabular}{c|c|c|c|c|c|c|c|cc}
$\chi_{14,1}$ & $\chi_{14,2}$ & $\chi_{24,1}$ & $\chi_{24,2}$ & $\chi_{31,1}$ & $\chi_{31,2}$ & \multicolumn{1}{c|}{$\chi_{32,1}$} & \multicolumn{1}{c|}{$\chi_{32,2}$} & \multicolumn{1}{c|}{$\chi_{43,1}$} & $\chi_{43,2}$ \\ \hline
$\epsilon_1$ & $-\epsilon_1$ & $\epsilon_2$ & $-\epsilon_2$ & $\epsilon_2$ & $-\epsilon_2$ & \multicolumn{1}{c|}{$\epsilon_1$} & \multicolumn{1}{c|}{$-\epsilon_1$} & \multicolumn{1}{c|}{$\epsilon_3$} & $-\epsilon_3$ \\ \hline\hline
$\Lambda_{12}$ & $\Lambda_{21}$ & $\Lambda_{34,1}$ & $\Lambda_{34,2}$ & $\Lambda_{34,3}$ & $\Lambda_{34,4}$ & $\chi_{01}$ & $\Lambda_{10}$ & \multicolumn{2}{c}{\multirow{2}{*}{}}                                    \\ \cline{1-8}
$-\epsilon_3$ & $-\epsilon_3$ & $\epsilon_1+\epsilon_2$ & $\epsilon_1+\epsilon_2$ & $\epsilon_1-\epsilon_2$ & $\epsilon_2-\epsilon_1$ & $v_1$ & $-v_2$ & \multicolumn{2}{c}{}
\end{tabular}\;.
\end{equation}
There are three equivariant parameters $\epsilon_{1,2,3}$. Together with the parameter $\epsilon_4=-\epsilon_1-\epsilon_2-\epsilon_3$ due to the Calabi-Yau condition, we have a 4d cyrstal \cite{Franco:2023tly,Bao:2024ygr}. The partition function reads
\begin{equation}
    \mathcal{Z}=\sum_{\mathscr{M}}\bm{q}^N \xi \, \mathcal{Z}_V \, \mathcal{Z}_\chi \, \mathcal{Z}_\Lambda \, \mathcal{Z}_\mathfrak{f}\;,
\end{equation}
where
\begin{subequations}
\begin{align}
    \mathcal{Z}_V=&\prod_{a,b\in Q_0}\prod_{\substack{\mathfrak{a}_i\in\mathscr{M}\\\mathfrak
    {a}_i\neq\mathfrak{a}_j\in\bigcap\limits_{\chi\in\{b\rightarrow a\}}\partial_{-\chi}\mathscr{M}}}E_{ij}(0)\;,\\
    \mathcal{Z}_\chi=&\prod_{a\in Q_0}\prod_{\mathfrak{a}_i\in\mathscr{M}}\prod_{\chi\in\{b\rightarrow a\}}\prod_{\substack{\mathfrak{a}_i\in\mathscr{M}\\\mathfrak{b}_j\in\partial_\chi\mathscr{M}}}\frac{1}{E_{ij}(-\epsilon_\chi)^{\delta_\chi(\mathfrak{b}_j)}}\;,\\
    \mathcal{Z}_\Lambda=&\prod_{a\in Q_0}\prod_{\mathfrak{a}_i\in\mathscr{M}}\prod_{\Lambda\in\{b\rightarrow a\}}\prod_{\substack{\mathfrak{a}_i\in\mathscr{M}\\\mathfrak{b}_j\in\partial_{-\cup\chi-\Lambda}\mathscr{M}}}E_{ij}(\epsilon_\Lambda)\;,\\
    \mathcal{Z}_\mathfrak{f}=&\frac{\prod\limits_{\substack{\mathfrak{a}_i\in\mathscr{M}\\(a=1)}}E_{00}(\epsilon(\mathfrak{a}_i)-v_2)}{\prod\limits_{\substack{\mathfrak{a}_i\in\mathscr{M}\backslash\mathscr{O}\\(a=1)}}E_{00}(\epsilon(\mathfrak{a}_i)-v_1)}\;.
\end{align}
\end{subequations}
Let us list some configurations at low levels. At level 1, there is one atom $\epsilon(\mathfrak{a})=v_1$ with $a=1$. We have
\begin{equation}
    \mathcal{Z}_1= \xi(1)E_{00}(\epsilon(\mathfrak{a})-v_2)\;.
\end{equation}
At level 2, there are two atoms\footnote{There is another configuration with two atoms where $\epsilon(\mathfrak{b})=v_1-\epsilon_1$. As this is similar, we shall omit it here.} $\epsilon(\mathfrak{a})=v_1$ and $\epsilon(\mathfrak{b})=v_1+\epsilon_1$ with $a=1$ and $b=4$. We have
\begin{equation}
    \mathcal{Z}_2=\xi(2)\frac{E_{00}(\epsilon(\mathfrak{a})-v_2)}{E_{\mathfrak{b},\mathfrak{a}}(\epsilon_1)}\;.
\end{equation}
At level 3, one possible configuration has $\epsilon(\mathfrak{a})=v_1$, $\epsilon(\mathfrak{b}_1)=v_1+\epsilon_1$ and $\epsilon(\mathfrak{b}_2)=v_1-\epsilon_1$ with $a=1$ and $b=4$. We have
\begin{equation}
    \mathcal{Z}_3=\xi(3)E_{00}(\epsilon(\mathfrak{a})-v_2)\;.
\end{equation}
Another possible configuration has\footnote{Again, there are three similar configurations with $\epsilon_1\rightarrow-\epsilon_1$ and/or $\epsilon_3\rightarrow-\epsilon_3$.} $\epsilon(\mathfrak{a})=v_1$, $\epsilon(\mathfrak{b})=v_1+\epsilon_1$ and $\epsilon(\mathfrak{c})=v_1+\epsilon_1+\epsilon_3$ with $a=1$, $b=4$ and $c=3$. We have
\begin{align}
    \mathcal{Z}_3&=\xi(3)\frac{E_{00}(\epsilon(\mathfrak{a})-v_2)E_{\mathfrak{c},\mathfrak{b}}(\epsilon_1+\epsilon_2)E_{\mathfrak{c},\mathfrak{b}}(\epsilon_1+\epsilon_2)E_{\mathfrak{c},\mathfrak{b}}(\epsilon_1-\epsilon_2)E_{\mathfrak{c},\mathfrak{b}}(\epsilon_2-\epsilon_1)}{E_{\mathfrak{b},\mathfrak{a}}(\epsilon_1)E_{\mathfrak{a},\mathfrak{c}}(-\epsilon_2)E_{\mathfrak{a},\mathfrak{c}}(\epsilon_2)E_{\mathfrak{c},\mathfrak{a}}(\epsilon_3)}\nonumber\\
    &=\xi(3)\frac{E_{00}(\epsilon(\mathfrak{a})-v_2)E_{\mathfrak{c},\mathfrak{b}}(\epsilon_1+\epsilon_2)E_{\mathfrak{c},\mathfrak{b}}(\epsilon_2-\epsilon_1)}{E_{\mathfrak{b},\mathfrak{a}}(\epsilon_1)E_{\mathfrak{c},\mathfrak{a}}(\epsilon_3)}\;.
\end{align}

\section{Emergent Limit Shapes}\label{limitshapes}

Having derived a statistical mechanical model of crystal melting, a natural question is to analyse its thermodynamic limit, where the fugacities are simultaneously scaled to zero. In the thermodynamic limit, we expect a smooth profile of crystals to emerge, and we can ask if such a profile (known as the limit shape) is related in any way to the geometries underlying the quiver gauge theory we started with. In fact, such a relation was anticipated in \cite{Okounkov:2003sp,Ooguri:2009ijd}, where atoms in the crystals are identified as the ``atoms of spacetime'' in quantum gravity---the discrete atoms in the crystals are regarded as quantized geometry, and conversely, a smooth classical geometry emerges in the thermodynamic limit of the model. 

Such a question was famously analysed for the Nekrasov partition function associated with four-dimensional $\mathcal{N}=2$ theories, where the limit shape reproduced the classical Seiberg-Witten geometry \cite{Nekrasov:2003rj}.

Another important insight comes from the analysis of the 
quiver gauge theories associated with toric Calabi-Yau three-folds \cite{Ooguri:2009ri} (see also \cite{cohn2001variational,Okounkov:2003sp,Kenyon:2003uj,Iqbal:2003ds,Yamazaki:2012cp}). More concretely, the thermodynamic limit shape of a molten crystal is governed by the Ronkin function \cite{Ronkin,PassareRullgard}
\begin{equation}
    R(x_1,x_2):=\frac{1}{(2\pi i)^2}\int_{\mathbb{T}^2}\log|P(\text{e}^{x_1}z_1,\text{e}^{x_2}z_2)|\frac{\d z_1}{z_1}\frac{\d z_2}{z_2}
\end{equation}
of the Newton polynomial $P(z_1,z_2)$. The BPS partition function in the thermodynamic limit can then be expressed using the Ronkin function as
\begin{equation}
    \mathcal{Z}\sim\exp\left(\frac{4}{g_s}\int_{\mathbb{R}^2}\d x_1\,\d x_2\,R(x_1,x_2)\right)\;,
\end{equation}
where $g_s$ denotes the fugacity associated with the D0-brane, and the BPS partition function is normalised to be 1 for the unmolten crystal to avoid divergence of the integral. This is done by subtracting the contributions from ``solid phases'', which will be explained below.

It is therefore natural to ask if there exists a suitable generalisation for more general two-supercharge theories discussed above---in the thermodynamic limit of the molten crystals, what is the limit shape that determines the profiles of the molten crystals, and how is this shape related to the geometry underlying the gauge theory we started with?

While this is an interesting question, this seems to be a challenging problem in general since our partition function \eqref{eq:schematic}, 
in particular the weight $w(\mathscr{M}; \bm{q}, \bm{\epsilon})$ associated with a molecule $\mathscr{M}$, is very complicated. Our strategy here is to concentrate on quiver gauge theories associated with toric Calabi-Yau four-folds, where we have a clear-cut geometry/combinatorics as a starting point. We then propose to replace the weight 
by a simpler function, as determined by the standard dimer partition function, as will be explained below. While this procedure will change the partition function in general, the procedure is known to give the precise BPS partition function for quivers associated with Calabi-Yau three-folds \cite{cohn2001variational,Okounkov:2003sp,Kenyon:2003uj,Ooguri:2009ri}, and 
hence we may hope our analysis will have some bearing to the BPS counting problem, as we will discuss further around \eqref{eq:refined_Z}.
In any case, a cautious reader can think of the analysis below as the 
discussion of the thermodynamic limit of the partition functions associated with three-dimensional dimer models.

\subsection{Newton Polynomials and Brane Brick Models}\label{Newtonpoly}
To discuss the limit shapes, we would first like to get the Newton polynomials from the brane brick models. Unlike the dimer models, there is no analogue of the Kasteleyn matrices \cite{Kasteleyn}. Nevertheless, given a brane brick model, the corresponding Newton polytope $\Delta$ is known. From the Newton polytope, we may write down the Newton polynomial in a way similar to the dimer case in \cite{cohn2001variational,Kenyon:2003uj} as
\begin{equation}
    P(z_1,z_2,z_3)=\sum_{M\in\mathcal{M}(G_1)}\text{e}^{-\mathcal{E}(M)}(-1)^{\sum\limits_{i=1}^3\sigma_i}z_1^{h_1}z_2^{h_2}z_3^{h_3}\;,\label{Pz1z2z3}
\end{equation}
where each lattice point $(h_1,h_2,h_3)\in\Delta\cap\mathbb{Z}^3$ corresponds to a monomial term with $z_1^{h_1}z_2^{h_2}z_3^{h_3}$, and $\sigma_i$ denotes the $i^\text{th}$ elementary symmetric sum of $h_1$, $h_2$, $h_3$. The notations in this expression would need some explanations.

Given a brane brick model, it can be viewed as a $\mathbb{Z}_3$-periodic graph $G$. Consider the graph $G_n$ as the quotient of $G$ by the translation action of $n\mathbb{Z}^3$. It is then a finite graph in the 3-torus $\mathbb{T}^3$. In particular, $G_1$ gives the fundamental domain of the brane brick model.

We refer the readers to \cite{Franco:2015tna,Franco:2015tya,Franco:2016nwv,Franco:2016qxh,Franco:2017cjj,Franco:2022iap,Franco:2022gvl,Franco:2022isw,Franco:2024lxs} for more details on the brane brick models for the 2d $\mathcal{N}=(0,2)$ quiver gauge theories constructed from toric CY fourfolds. We shall only mention the relevant concepts involved in the discussions here.

\begin{figure}[ht]
    \centering
    \includegraphics[width=10cm]{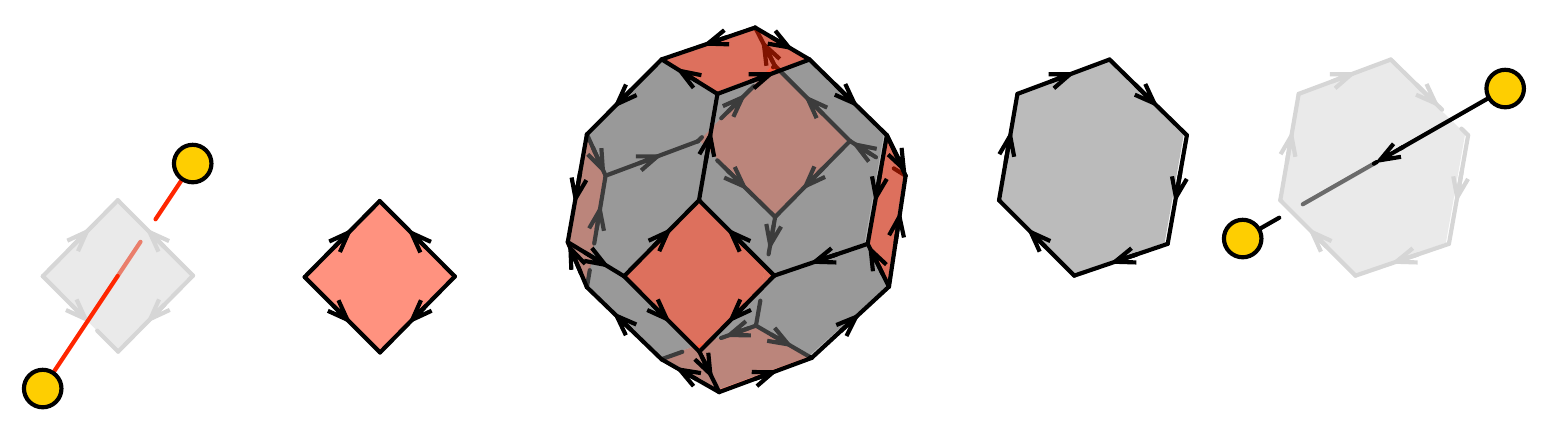}
    \caption{The brane brick model is the dual graph of the quiver. Each brick corresponds to a gauge node, and the faces are the arrows in the quiver. The oriented (resp.~unoriented) faces give the chiral (resp.~Fermi) multiplet, which are oriented (resp.~unoriented) arrows in the quiver. Figure taken from \cite{Franco:2015tya}.}\label{brickorientation}
\end{figure}

The faces in the brane brick model can be either oriented or unoriented. The oriented (resp.~unoriented) faces correspond to the chiral (resp.~Fermi) multiplets. An illustration can be found in \cref{brickorientation}.
For the 2d $\mathcal{N}=(0,2)$ gauge theory, it is equivalent to choose either the $J$-term interaction $\Lambda\chi_{a_1}\dots\chi_{a_k}-\Lambda\chi_{b_1}\dots\chi_{b_l}$ or the $E$-term interaction $\overline{\Lambda}\chi_{c_1}\dots\chi_{c_m}-\overline{\Lambda}\chi_{d_1}\dots\chi_{d_n}$. Notice that in a toric theory, each Fermi multiplet would always appear twice in the interaction so that the chiral multiplets together with the Fermi would form two plaquettes in the periodic quiver, which in turn corresponds to an edge in the brane brick model. For instance, if we choose the $J$-terms, then the two plaquettes are given by $\Lambda\chi_{a_1}\dots\chi_{a_k}$ and $\Lambda\chi_{b_1}\dots\chi_{b_l}$. In a purely combinatorial way, it is sufficient to know that
\begin{itemize}
    \item each unoriented arrow is incident to four plaquettes in the quiver,
    \item and the four plaquettes are divided into two pairs called the $J$-term plaquettes and the $E$-term plaquettes.
\end{itemize}

A brick matching \cite{Franco:2015tna,Franco:2015tya} of the brane brick model is a collection of faces that can be combinatorially defined as follows. For each Fermi pair $\Lambda$ and its conjugate $\overline{\Lambda}$, we choose the chiral multiplets such that they would cover either the $J$-term plaquettes or the $E$-term plaquettes. Moreover, for each single plaquette in each chosen pair, there is only one such chosen chiral. If the $J$-term (resp.~$E$-term) plaquettes are chosen for a Fermi pair, then $\overline{\Lambda}$ (resp.~$\Lambda$) is chosen in the brick matching. For example, the quiver for the $\mathbb{C}^4$ case has one single gauge node and seven loops on the node, four (resp.~three) of which are chiral (resp.~Fermi) multiplets. The brane brick model is the tiling of the truncated octahedron in \cref{brickorientation}:
\begin{equation}
    \includegraphics[width=5cm]{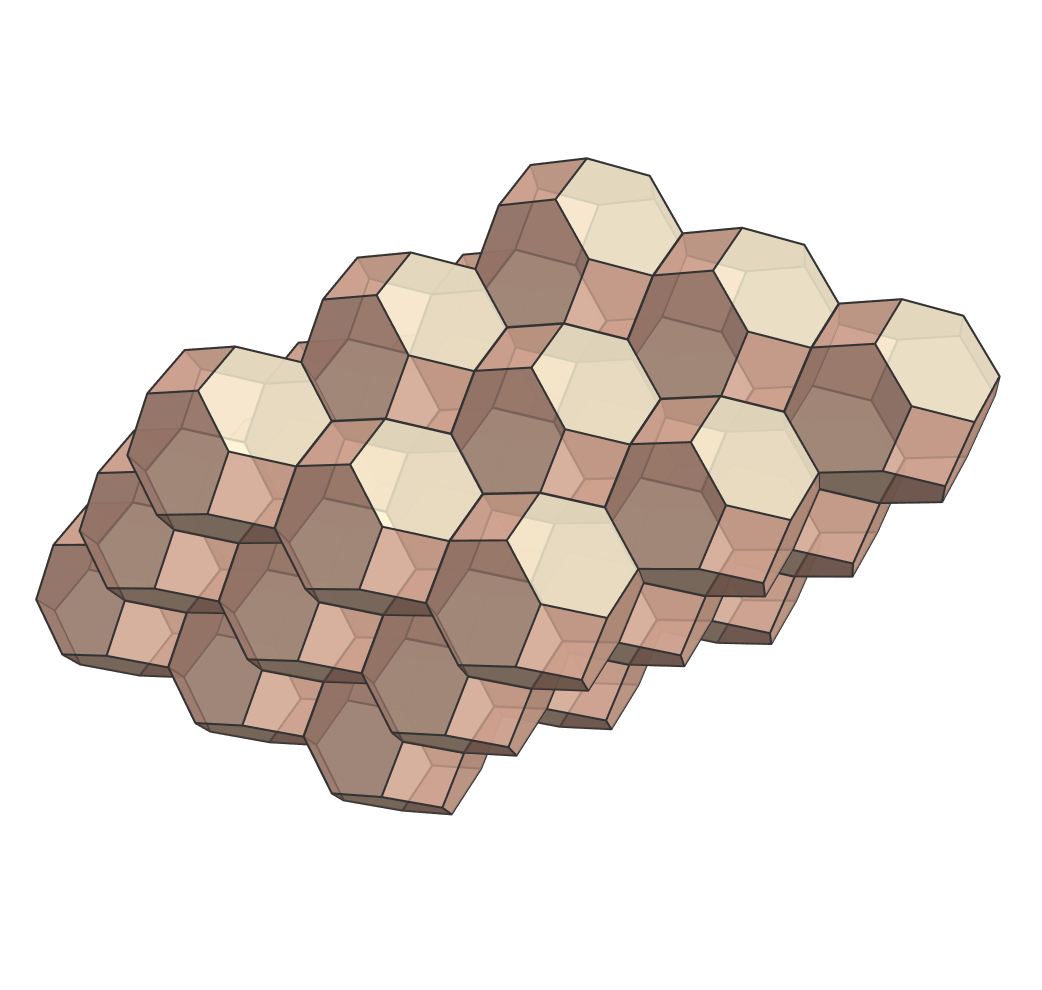}\;.
\end{equation}
The $J$-/$E$-terms are
\begin{equation}
    \begin{tabular}{cc}
             $J$ & $E$ \\
             $\Lambda_1(YZ-ZY)$ & $\overline{\Lambda}_1(DX-XD)$ \\
             $\Lambda_2(ZX-XZ)$ & $\overline{\Lambda}_2(DY-YD)$ \\
             $\Lambda_3(XY-YX)$ & $\overline{\Lambda}_3(DZ-ZD)$
    \end{tabular}\;.
\end{equation}
Here, $X$, $Y$, $Z$, $D$ are the chiral multiplets, and $\Lambda_i$ are the Fermi multiplets. There are four possible brick matchings, $M_1=\left\{X,\Lambda_1,\overline{\Lambda}_2,\overline{\Lambda}_3\right\}$, $M_2=\left\{Y,\overline{\Lambda}_1,\Lambda_2,\overline{\Lambda}_3\right\}$, $M_3=\left\{Z,\overline{\Lambda}_1,\overline{\Lambda}_2,\Lambda_3\right\}$, $M_4=\left\{D,\Lambda_1,\Lambda_2,\Lambda_3\right\}$.

We notice that the unoriented faces are then always in the brick matching, and the difference is whether $\Lambda$ or its conjugate $\overline{\Lambda}$ is chosen. Nevertheless, for our purpose here, it suffices to consider the chiral multiplets. Therefore, combinatorially, we are just choosing the oriented faces in the brane brick model such that
\begin{itemize}
    \item for each plaquette quadruple, only the $J$-term plaquette pair or the $E$-term plaquette pair is chosen,
    \item and precisely one oriented arrow would appear in each chosen plaquette.
\end{itemize}
The information of the brick matchings can be summarised in the brick matching matrix with rows (resp.~columns) being the chiral and Fermi multiplets (resp.~brick matchings). The entry is 1 if the field is in the brick matching and 0 otherwise. In the above example, the brick matching matrix is
\begin{equation}
    P=\begin{pmatrix}
\begin{array}{c|cccc}
 & M_1 & M_2 & M_3 & M_4 \\ \hline
X & 1 & 0 & 0 & 0 \\
Y & 0 & 1 & 0 & 0 \\
Z & 0 & 0 & 1 & 0 \\
D & 0 & 0 & 0 & 1 \\
\end{array}
\end{pmatrix}\;.
\end{equation}
We have omitted the Fermi multiplets $\Lambda_i$ and $\overline{\Lambda}_i$ in the brick matching matrix since we shall only need to consider the chiral multiplets here.

Given a graph $G_n$, the set of its brick matchings is denoted as $\mathcal{M}(G_n)$. For the theory associated to the brane brick model with fundamental domain $G_1$, the brick matchings $\mathcal{M}(G_n)$ should be regarded as a combinatorial object. In other words, choosing $\Lambda$ and the $E$-term plaquettes (or $\overline{\Lambda}$ and the $J$-term plaquettes) in $G_1$ does not necessarily imply the same choice in the remaining $n^2-1$ copies. Alternatively, we can think of $\mathcal{M}(G_n)$ as the set of brick matchings for the theory associated to the Calabi-Yau fourfold $X/(\mathbb{Z}_n\times\mathbb{Z}_n\times\mathbb{Z}_n)$, where $G_1$ is the fundamental domain of the theory associated to the Calabi-Yau fourfold $X$. In other words\footnote{In \cite{Pomoni:2023nlf}, the authors discussed how higher rank DT countings can be related to rank-1 counting via the orbifolds of the geometry (see also \cite{Fasola:2023ypx}). How $G_n(X)$ could be used to count the DT4 invariants in the orbifold cases has not yet been worked out, and it is curious to see the precise correspondence between the ranks and the orbifolds here. We would like to thank the referee for pointing this out.}, $G_1(X/(\mathbb{Z}_n\times\mathbb{Z}_n\times\mathbb{Z}_n))=G_n(X)$.

Now, for each face $\chi$ in the graph\footnote{Since we only need to consider the chiral multiplets, we shall also use $\chi$ to denote the corresponding face in the brane brick model}, we assign a weight $\text{e}^{-\mathcal{E}(\chi)}$ to it, where $\mathcal{E}(\chi)$ can be taken as any arbitrary real number called the energy of the face. The energy of a brick matching $M$ is then given by\footnote{Here, we have not included any weights from the Fermi multiplets. Although it is not clear whether there exists a 3d analogue of the Kasteleyn matrix that determines the coefficients in the Newton polynomial in a canonical way, that the Fermi multiplets could be some ``redundancies'' might be expected from various perspectives. The most direct way to see that we are not including them is as follows. Later, we will consider certain products of the Newton polynomials of $G_1$, and the energies of the Fermi multiplets should not be mixed with those of the chiral multiplets as the products are expected to be related to the Newton polynomials of $G_n$. Alternatively, we can take $\mathcal{E}(\Lambda)=\mathcal{E}\left(\overline{\Lambda}\right)$. As the unoriented faces always appear in all the brick matchings, they would give an overall factor in the Newton polynomial regardless of the choice of the Fermi multiplets or their conjugates.} $\mathcal{E}(M)=\sum\limits_{\chi\in M}\mathcal{E}(\chi)$, and the weight of $M$ is $\text{e}^{-\mathcal{E}(M)}$.

To get the Newton polynomial in \eqref{Pz1z2z3}, we need to assign the brick matchings to the lattice points of the Newton polygon. In fact, this can be determined using the brick matching matrix. In short, one obtains the $J$-/$E$-term matrix $Q_{JE}$ by taking the kernel of the brick matching matrix, that is, $Q_{JE}=\ker(P)$. Moreover, the $D$-term matrix $Q_D$ is obtained from $d=Q_D\cdot P^\text{T}$, where $d$ is the incidence matrix of the quiver\footnote{The incidence matrix has rows (resp.~columns) indicating the quiver nodes (resp.~chiral multiplets). If the node is the source (resp.~target) of the arrow, then the corresponding entry is $+1$ (resp.~$-1$). Otherwise, the entry is $0$.}. They capture the charges of the brick matchings (i.e., GLSM fields) under the $J$-/$E$-term and $D$-term relations. Consider the total charge matrix by combining the two matrices, namely, $Q_t=\binom{Q_{JE}}{Q_D}$. The lattice points corresponding to the brick matchings are encoded by the kernel $G_t=\ker(Q_t)$, where each column is a brick matching, and the four rows in each column give the coordinates of the lattice points\footnote{Recall that the toric Calabi-Yau fourfold is represented by a cone whose base is co-hyperplanar, and the base is the Newton polytope. Therefore, there is always one row in $G_t$ being $(1,1,\dots,1)$, and the remaining three rows indicate the coordinates of the lattice points in the Newton polytopes.}. See \cite{Franco:2015tna} for more details.

The above method to get the correspondence between the matchings and the lattice points is analogous to the story in dimer models \cite{Feng:2000mi}. For dimer models, there is another method that would lead to the same result. One can take the ``magnetically altered'' Kasteleyn matrix whose determinant gives the Newton polynomial \cite{Kenyon:2003uj}. Here, we do not have an analogue of the Kasteleyn matrix, but we may still propose a similar method to obtain the Newton polynomial by turning on some ``magnetic fluxes'' through the faces in the brane brick model. More concretely, we take the three oriented paths $\gamma_{1,2,3}$ along the three cycles of the 3-torus $\mathbb{T}^3$. We pick a reference matching and change it to another matching. If a face of the same orientation as the path $\gamma_i$ appears (resp.~disappears) when we change the reference matching to the new matching, then we increase (resp.~decrease) the height $h_i$ by 1. If a face of the opposite orientation as the path $\gamma_i$ appears (resp.~disappears) when we change the reference matching to the new matching, then we decrease (resp.~increase) the height $h_i$ by 1. Here, we say that the face is of the same orientation of the path if their orientations satisfy the right-hand rule. Equivalently, this means that the corresponding chiral has the same orientation as the ``magnetic flux''.

\paragraph{Example 1} Let us illustrate this with some examples. Consider the $\mathbb{C}^4$ case whose brane brick model and brick matching matrix are given as above. Compared to the brane bricks, it would be easier to visualise this using the periodic quiver:
\begin{equation}
    \includegraphics[width=5cm]{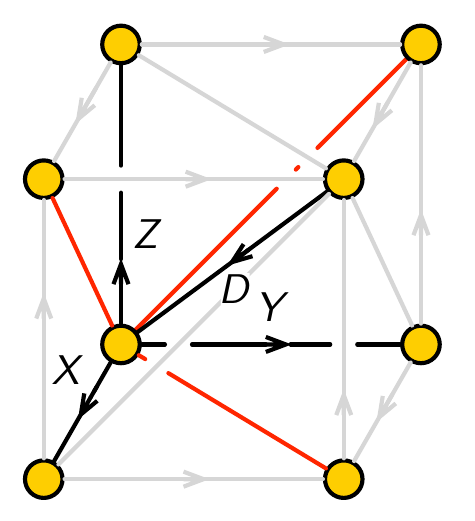}\;.
\end{equation}
We shall choose the three paths to align with the directions of $X$, $Y$ and $Z$ respectively so that each path would only pass one face. The reference brick matching is taken to be $p_4=\{D\}$ (where we have omitted the Fermi multiplets). Then it is clear that the height changes are
\begin{align}
    &(h_1,h_2,h_3)_{M_1}=(1,0,0)\;,\quad(h_1,h_2,h_3)_{M_2}=(0,1,0)\;,\nonumber\\
    &(h_1,h_2,h_3)_{M_3}=(0,0,1)\;,\quad(h_1,h_2,h_3)_{M_4}=(0,0,0)\;.
\end{align}
Therefore, the Newton polynomial is
\begin{equation}
    P(z_1,z_2,z_3)=-\text{e}^{-\mathcal{E}(X)}z_1-\text{e}^{-\mathcal{E}(Y)}z_2-\text{e}^{-\mathcal{E}(Z)}z_3+\text{e}^{-\mathcal{E}(D)}\;.
\end{equation}

\paragraph{Example 2} As another example, consider the $Q^{1,1,1}$ case whose toric diagram is
\begin{equation}
    \includegraphics[width=5cm]{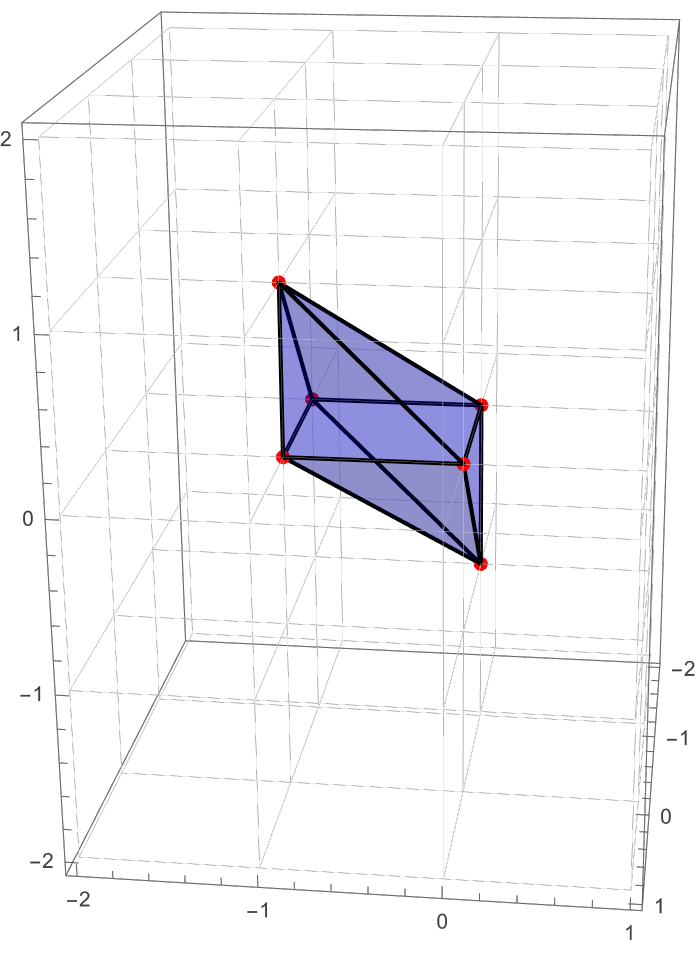}\;.
\end{equation}
The periodic quiver is
\begin{equation}
    \includegraphics[width=5cm]{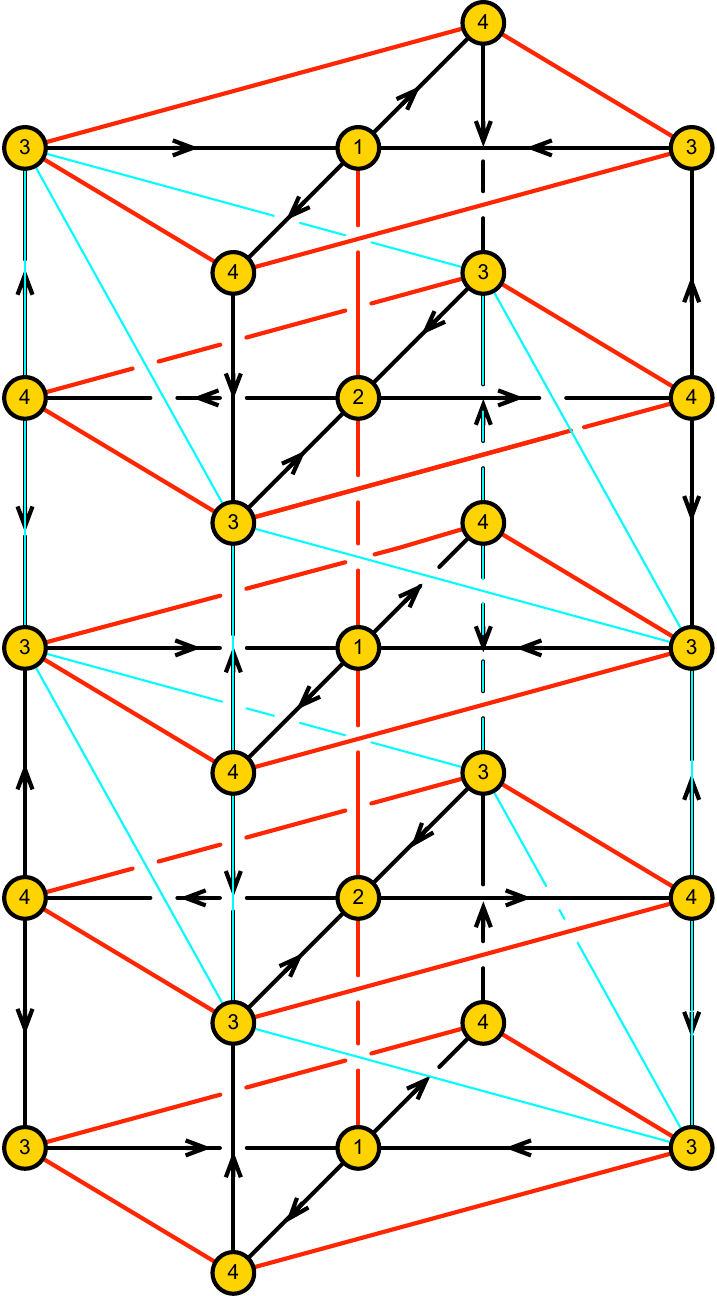}\;,
\end{equation}
where the cyan lines indicate the fundamental domain. The brick matching matrix is given by
\begin{equation}
    P=\begin{pmatrix}
\begin{array}{c|cccccc}
 & M_1 & M_2 & M_3 & M_4 & M_5 & M_6 \\ \hline
X^+_{14} & 1 & 0 & 0 & 0 & 0 & 0 \\
X^-_{14} & 0 & 1 & 0 & 0 & 0 & 0 \\
X^+_{32} & 1 & 0 & 0 & 0 & 0 & 0 \\
X^-_{32} & 0 & 1 & 0 & 0 & 0 & 0 \\
X^+_{24} & 0 & 0 & 1 & 0 & 0 & 0 \\
X^-_{24} & 0 & 0 & 0 & 1 & 0 & 0 \\
X^+_{31} & 0 & 0 & 1 & 0 & 0 & 0 \\
X^-_{31} & 0 & 0 & 0 & 1 & 0 & 0 \\
X^+_{43} & 0 & 0 & 0 & 0 & 1 & 0 \\
X^-_{43} & 0 & 0 & 0 & 0 & 0 & 1 \\
\end{array}
\end{pmatrix}\;.
\end{equation}
The $+$ superscripts are for the directions pointing outwards, rightwards, and downwards. The paths are given as follows: $\gamma_1$ passes through the faces $+X^+_{14}$, $+X^-_{31}$, $+X^-_{43}$; $\gamma_2$ passes through the faces $+X^+_{14}$, $+X^+_{31}$, $+X^+_{43}$; $\gamma_3$ passes through the faces $-X^+_{43}$, $+X^-_{43}$. Here, the signs in front of the faces/chiral multiplets indicate whether the faces are in the same ($+$) or opposite ($-$) directions of the paths, and we have omitted the paths in the above figure to avoid clutter. Choose $M_1=\left\{X^+_{14},X^+_{32}\right\}$ to be the reference brick matching. Then the Newton polynomial is
\begin{equation}
    P(z_1,z_2,z_3)=\text{e}^{-\mathcal{E}(M_1)}-\text{e}^{-\mathcal{E}(M_2)}z_1^{-1}z_2^{-1}-\text{e}^{-\mathcal{E}(M_3)}z_1^{-1}-\text{e}^{-\mathcal{E}(M_4)}z_2^{-1}-\text{e}^{-\mathcal{E}(M_5)}z_1^{-1}z_3^{-1}-\text{e}^{-\mathcal{E}(M_6)}z_2^{-1}z_3\;.
\end{equation}

\subsection{The Thermodynamic Limit}\label{thermolim}
Let us now consider the profiles of the molten crystals in the thermodynamic limit. We shall mainly follow the strategy for the dimer models in \cite{Kenyon:2003uj} and try to generalise this to the brane brick models.

On $\mathcal{M}(G_n)$, we may define a probability measure $\mu_n(M)={\text{e}^{-\mathcal{E}(M)}}/{Z(G_n)}$ for any brick matching $M\in\mathcal{M}(G_n)$. The normalising constant
\begin{equation}
    Z(G_n)=\sum_{M\in\mathcal{M}(G_n)}\text{e}^{-\mathcal{E}(M)}
\end{equation}
is called the partition function of the graph/brane brick model $G_n$ (not to be confused with the BPS partition function).

For $G_1$, the partition function of the brane brick model can be written as
\begin{align}
    Z(G_1)=\frac{1}{4}&(-3P(1,1,1)+P(-1,1,1)+P(1,-1,1)+P(1,1,-1)\nonumber\\
    &+P(-1,-1,1)+P(-1,1,-1)+P(1,-1,-1)+P(-1,-1,-1))\;.
\end{align}
For $G_n$, we have the similar expression with $P_1(z)=P(z)$ replaced by some Newton polynomial $P_n(z)$ (where we have collectively denote $z_{1,2,3}$ as $z$). This Newton polynomial $P_n(z)$ is actually the Newton polynomial for the quotient Calabi-Yau fourfold as discussed above.

For the dimer models, $P_n(z_1,z_2)$ can be expressed as
\begin{equation}
    P_n(z_1,z_2)=\prod_{w_1^n=z_1}\prod_{w_2^n=z_2}P(w_1,w_2)
\end{equation}
since the Kastelyn matrix is block diagonal (and the Newton polynomial is its determinant). Due to the lack of a similar concept for the brane brick models, it is not clear whether this product expression would still hold for $P_n(z)$. Therefore, let us write it as
\begin{equation}
    P_n(z)=C_n(z)Q_n(z)\;,\quad Q_n(z)=\prod_{w_1^n=z_1^n}\prod_{w_2^n=z_2^n}\prod_{w_3^n=z_3^n}P(z)\;,
\end{equation}
where $C_n(z)$ is some rational function. Then the partition function of $G_n$ may be written as
\begin{equation}
    Z(G_n)=\frac{1}{4}\left(-3Z^{(000)}_n+Z^{(100)}_n+Z^{(010)}_n+Z^{(001)}_n+Z^{(110)}_n+Z^{101}_n+Z^{(011)}_n+Z^{(111)}_n\right)\;,
\end{equation}
where
\begin{equation}
    Z^{(\theta)}_n=Z^{(\theta_1\theta_2\theta_3)}_n=P_n\left((-1)^{\theta_1},(-1)^{\theta_2},(-1)^{\theta_3}\right)=C_n\left((-1)^{\theta}\right)Q_n\left((-1)^\theta\right)\;.
\end{equation}

Consider
\begin{equation}
    \log Z^{(\theta)}_n=\log C_n\left((-1)^{\theta}\right)+\log Q_n\left((-1)^{\theta}\right)\;.
\end{equation}
In the $n\rightarrow\infty$ limit, the Riemann sum part $\log Q_n$ can be written as an integral. In other words\footnote{We have assumed that the integral would approximate the Riemann sum and that the limit would exist. See \cite{Kenyon:2003uj} for more details.},
\begin{equation}
    \lim_{n\rightarrow\infty}\frac{1}{n^3}\log Z^{(\theta)}_n=\lim_{n\rightarrow\infty}\frac{1}{n^3}\log C_n\left((-1)^{\theta}\right)+\frac{1}{(2\pi i)^3}\int_{\mathbb{T}^3}\log|P(z)|\frac{\d z_1}{z_1}\frac{\d z_2}{z_2}\frac{\d z_3}{z_3}\;.
\end{equation}
As $Z^{(\theta)}_n\leq Z(G_n)\leq4\max\limits_{\theta}\left\{Z^{(\theta)}_n\right\}$, we have
\begin{equation}
    \log Z:=\lim_{n\rightarrow\infty}\frac{1}{n^3}\log Z(G_n)=\lim_{n\rightarrow\infty}\frac{1}{n^3}\log C_n\left((-1)^{\theta}\right)+\frac{1}{(2\pi i)^3}\int_{\mathbb{T}^3}\log|P(z)|\frac{\d z_1}{z_1}\frac{\d z_2}{z_2}\frac{\d z_3}{z_3}\;,
\end{equation}
where $Z$ is called the partition function of the brane brick model per fundamental domain. Besides the Newton polynomial $P(z)$, there is another part with $C_n$ in the expression. We shall now show that this contribution would vanish when $n\rightarrow\infty$.

By definition, $C_n(z)=P_n(z)/Q_n(z)$, where $P_n(z)$ and $Q_n(z)$ have the same monomial terms but different (non-zero) coefficients. Write $P_n(z)=\sum\limits_va_vz^v$ and $Q_n(z)=\sum\limits_vb_vz^v$ with $v$ denoting the coordinates of the lattice points. Then $C_n\left((-1)^{\theta}\right)=\left(\sum\limits_v(\pm a_v)\right)\bigg/\left(\sum\limits_v(\pm b_v)\right)$. Similar to the condition for the Riemann sum to be approximated by the integral, we require $\left((-1)^{\theta_1},(-1)^{\theta_2},(-1)^{\theta_3}\right)$ to be away from the zeros of $Q_n$ (at least for most $n$). Now, when one considers the $n^3$ copies of the fundamental domain, the number of brick matchings would scale as $\sim\text{e}^{n^3}$. Therefore, the numerator and the denominator of $C_n$ would scale as $\sim A^{n^3}$ and $\sim B^{n^3}$ respectively. As a result, $\lim\limits_{n\rightarrow\infty}\frac{1}{n^3}\log C_n\left((-1)^{\theta}\right)=0$. The partition function of the brane brick model per fundamental domain is then the Mahler measure of the Newton polynomial:
\begin{equation}
    \log Z=\frac{1}{(2\pi i)^3}\int_{\mathbb{T}^3}\log|P(z)|\frac{\d z_1}{z_1}\frac{\d z_2}{z_2}\frac{\d z_3}{z_3}\;.
\end{equation}
As we can see, although the Newton polynomial $P_n(z)$ for the $n^3$ copies of the fundamental domain may or may not coincide with the above products of the Newton polynomial $P(z)$, the partition function of the brane brick model per fundamental domain is still determined by the Newton polynomial $P(z)$.

More generally, we can turn on the magnetic fluxes giving $\left(\text{e}^{x_1}z_1\right)^{h_1}\left(\text{e}^{x_2}z_2\right)^{h_2}\left(\text{e}^{x_3}z_3\right)^{h_3}$ in the Newton polynomial, where $x_{1,2,3}\in\mathbb{R}$. The logarithm of the partition function of the brane brick model per fundamental domain is then given by the Ronkin function $R(x)$ of the Newton polynomial:
\begin{equation}
    \log Z(x)=R(x)=\frac{1}{(2\pi i)^3}\int_{\mathbb{T}^3}\log|P(\text{e}^xz)|\frac{\d z_1}{z_1}\frac{\d z_2}{z_2}\frac{\d z_3}{z_3}\;.
\end{equation}
As a 3d analogue of the surface tension for the dimer model in \cite{Kenyon:2003uj}, we may also write the free energy per fundamental domain as the Legendre transform of $Z(x)$, that is,
\begin{equation}
    \sigma(s)=\max_{(x_1,x_2,x_3)}(-Z(x)+x_1s_1+x_2s_2+x_3s_3)\;.
\end{equation}

Having seen that the partition function of the brane brick model per fundamental domain is given by the Ronkin function, we are now going to discuss the BPS partition function. Similar to the discussions in \cite{Ooguri:2009ri}, we may consider the height function $\mathfrak{h}$ in the brane brick model defined as follows. Starting with a reference brick matching, for any other brick matching, consider their superposition. The height function is 0 far away from the region where the two brick matchings differ, and it increases by 1 every time one crosses a closed face given by the overlap of the two brick matchings as one moves inside. In the large $n$ limit, we may consider the expression
\begin{equation}
    \label{eq:refined_Z}
    \mathscr{Z}
    \sim\exp\left(n^3\max_{\mathfrak{h}}\int_{(0,1)^3}\d x_1\,\d x_2\,\d x_3\,(-\sigma(\nabla\mathfrak{h})-g_sn\mathfrak{h}(x_1,x_2,x_3))\right)\;.
\end{equation}
As the Ronkin function is the Legendre transform of the free energy per fundamental domain, this becomes
\begin{equation}
    \mathscr{Z}\sim\exp\left(n^3\int_{(0,1)^3}\d x_1\,\d x_2\,\d x_3\,R\left(\frac{g_sn}{2}x_1,\frac{g_sn}{2}x_2,\frac{g_sn}{2}x_3\right)\right)\;.
\end{equation}
Rescaling the integration variables by $g_sn/2$, we have
\begin{equation}
    \mathscr{Z}\sim\exp\left(\frac{8}{g_s^3}\int_{\mathbb{R}^3}\d x_1\,\d x_2\,\d x_3\,R(x)\right)
\end{equation}
in the thermodynamic limit $n\rightarrow\infty$. This integral is, of course, divergent. Therefore, we shall normalise it by setting one of the empty configurations to be 1. This is equivalent to subtracting the unbounded linear parts of the Ronkin function so that the integral becomes finite. The unbounded linear parts are known to be in the ``solid phases'' as discussed below.

However, as we have seen above, the BPS partition function $\mathcal{Z}$ would contain non-trivial information in terms of the equivariant weights. Therefore, we expect some extra pieces to be added to $\mathscr{Z}$ to give the actual BPS partition function:
\begin{equation}
    \mathcal{Z}\sim\exp\left(n^3\max_{\mathfrak{h}}\int_{(0,1)^3}\d x_1\,\d x_2\,\d x_3\,(-\sigma(\nabla\mathfrak{h})-\sigma_\text{equiv}(\nabla\mathfrak{h})-g_sn\mathfrak{h}(x_1,x_2,x_3))\right)\;.
\end{equation}
Here, we have denoted the extra contributions as $\sigma_\text{equiv}(\nabla\mathfrak{h})$. Despite this notation, it is not clear whether $\sigma_\text{equiv}(\nabla\mathfrak{h})$ may be regarded as some local ``fluctuations'' of the limit shape so that one can consider some modified free energy per fundamental domain $\sigma_\text{modified}(\nabla\mathfrak{h})=\sigma(\nabla\mathfrak{h})+\sigma_\text{equiv}(\nabla\mathfrak{h})$. One may hope that the actual BPS partition function would then be written as
\begin{equation}
    \mathcal{Z}\overset{?}{\sim}\exp\left(\frac{8}{g_s^3}\int_{\mathbb{R}^3}\d x_1\,\d x_2\,\d x_3\,R_\text{modified}(x)\right)\;,
\end{equation}
where $R_\text{modified}(x)$ is the Ronkin function of the same Newton polynomial $P(z)$ but with coefficients different from those in $\mathscr{Z}$ so as to contain the information of the non-trivial equivariant parameters. 

\begin{figure}[ht]
    \centering
    \includegraphics[width=5cm]{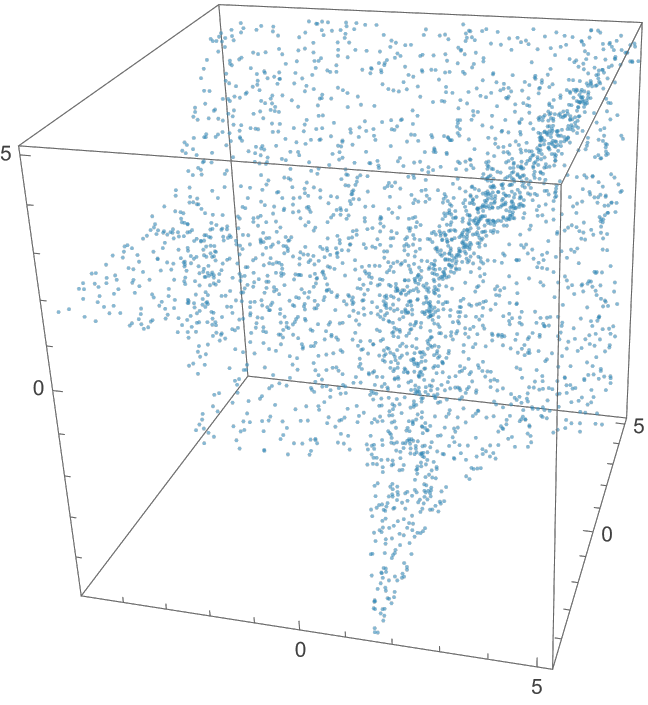}
    \includegraphics[width=5cm]{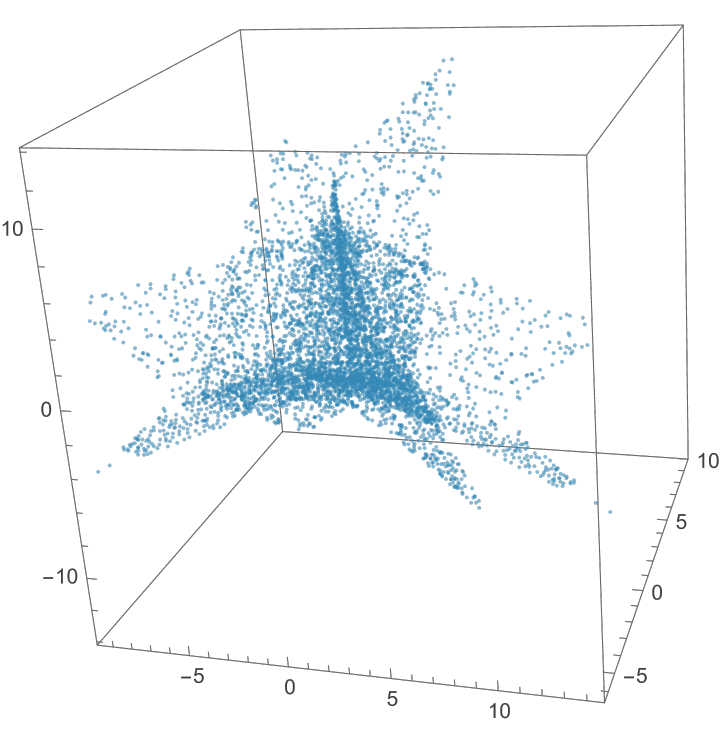}
    \includegraphics[width=5cm]{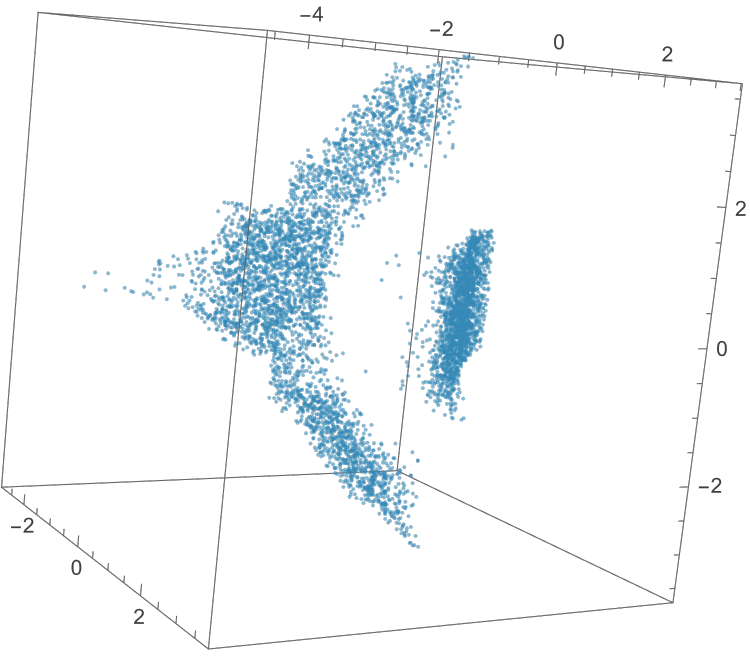}
    \caption{Some illustrations of the amoebae. Left: The amoeba for the $\mathbb{C}^4$ with the Newton polynomial $P(z)=z_1+z_2+z_3-5$. Middle: The amoeba for the so-called $K^{2,3}$ case whose Newton polynomial is $P(z)=z_1^{-1}+z_1z_2+z_2^{-1}z_3+z_2^{-1}z_3^{-1}+z_2^{-1}-10$. There is a gas phase inside. Right: To see the gas phase more clearly, we have zoomed in near the origin.}\label{amoebaex}
\end{figure}

Although the actual BPS partition function requires future study, the above discussions indicate that at least the limit shape of the molten crystal is given by the Ronkin function. In other words, the profile of the molten crystal in the limit shape is given by the profile of the Ronkin function called the amoeba:
\begin{equation}
    \mathcal{A}_{P(z)}=\left\{\left(\log|z_1|,\log|z_2|,\log|z_3|\right)\;|\;P(z_1,z_2,z_3)=0\right\}\;.
\end{equation}
Some illustrations of the amoebae can be found in \cref{amoebaex}.
In the Ronkin function, its different regions would be projected to three different cases in $\mathbb{R}^3$. If a region corresponds to the amoeba, then we say that it is the liquid phase. If it corresponds to a complementary region of the amoeba that is unbounded (resp.~bounded), then we say that it is a gas (resp.~solid) phase.

Computing the exact expressions of the Ronkin functions is notoriously hard. For instance, for the dimer models, only the $\mathbb{C}^3$ case $P(z_1,z_2)=az_1+bz_2+c$ \cite{maillot1997g} and the $F_0$ case $P(z_1,z_2)=z_1+z_1^{-1}+z_2+z_2^{-1}+4$ \cite{roy2024generalized} are known (see also, for example, \cite{vandervelde2003formula,Zahabi:2018znq} for relevant discussions on the Mahler measure, and \cite{de2021limit} for connections to integrable systems\footnote{We would like to thank Ali Zahabi for pointing out some references.}). Nevertheless, some of its properties have been well-studied \cite{forsberg2000laurent,PassareRullgard,mikhalkin2001amoebas}. Let us make some comments on the Ronkin functions and the discussions here:
\begin{itemize}
    \item The Ronkin function is strictly convex over $\mathcal{A}_{P(z)}$ and linear over each component of $\mathbb{R}^3\backslash\mathcal{A}_{P(z)}$. The gradient $\nabla$ of the Ronkin function satisfying $\text{Int}(\Delta)\subseteq\nabla R(x)\subseteq\Delta$, where $\text{Int}(\Delta)$ is the interior of the Newton polytope $\Delta$. In particular, for each component $E$ of $\mathbb{R}^3\backslash\mathcal{A}_{P(z)}$ (i.e., a solid phase or a gas phase), $\nabla R(E)=(a_1,a_2,a_3)$, where $(a_1,a_2,a_3)\in\Delta\cap\mathbb{Z}^3$ is the lattice point corresponding to the component $E$.
    \item The total mass defined as $\int_{\mathbb{R}^3}\det(\text{Hess}(R(x)))\d^3x$ is equal to the volume of the Newton polytope. In this sense, the ``fluctuation'' of the limit shape is bounded. Similar to the discussions in \cite{Nekrasov:2003rj} (see also, for example, \cite{Grekov:2023psy}), it could be possible that the Monge-Amp\'ere measure $\det(\text{Hess}(R(x)))\d^3x$ is related to the $\bm{Y}$-operators and $qq$-characters in the gauge origami system \cite{Nekrasov:2015wsu,Kimura:2015rgi,Kimura:2023bxy}.
    \item The Laurent coefficients of the Newton polynomial in the sense of \cite{forsberg2000laurent} may be viewed as a ``basis'' of (the entries of) the inverse Kasteleyn matrix in the dimer model case. Moreover, the matrix formed by the Laurent coefficients is block diagonal (although it is factorised in a way different from the Kasteleyn matrix). This might shed light on introducing some 3d analogue of the Kasteleyn matrix for the brane brick models.
    \item For the Nekrasov partition function, $\log\mathcal{Z}$ scales as $O(1/\epsilon^2)$ in the limit $\epsilon \to 0$ of the equivariant parameter $\epsilon$. Here we have the scaling $\log\mathcal{Z}\sim O(1/\epsilon^3)$, since we are in one dimension higher (in the sense of the statistical model and the Calabi-Yau geometry). It would be interesting to see whether and how this could give rise to some Seiberg-Witten-like superpotential, thereby generalising the classic Seiberg-Witten theory.
\end{itemize}

\section*{Acknowledgement}
We would like to thank Darius Dramburg and Jiaqun Jiang for enjoyable discussions. JB and MY are supported in part by the JSPS Grant-in-Aid for Scientific Research (Grant No.~23K25865). MY is also supported in part by JST, Japan (CREST Grant No.~JPMJCR26XA, Moonshot R\&D Grant No.~JPMJMS2061).

\appendix

\section{The Jeffrey-Kirwan Residue Formula}\label{JKres}
In this appendix, we shall recall the JK residue formula for the supersymmetric partition function:
\begin{equation}
	\mathcal{Z}(\bm{\epsilon})=\frac{1}{|\mathcal{W}|}\sum_{\bm{u}^*\in\mathfrak{M}^*_\text{sing}}\textrm{JK-Res}_{\bm{u}=\bm{u}^*}(\bm{\mathsf{Q}}(\bm{u}^*),\eta)\, Z_{\textrm{1-loop}}(\bm{\epsilon}, \bm{u}) \;.
\end{equation}
Let us now explain the ingredients in detail.

\paragraph{The space $\mathfrak{M}$} The space $\mathfrak{M}$ is defined to be $\mathfrak{h}_{\mathbb{C}}/\mathsf{Q}^{\vee}$, where $\mathfrak{h}$ is the Cartan subalgebra and $\mathsf{Q}^{\vee}$ is the coroot lattice. In the one-loop determinant, each multiplet gives rise to a hyperplane $H_i=\{\mathsf{Q}_i(u)+\dots=0\}\subset\mathfrak{M}$ with covector $\mathsf{Q}_i\in\mathfrak{h}^*$. The union of the hyperplanes is $\mathfrak{M}_\text{sing}=\bigcup\limits_iH_i$.

To compute the residues, we need to consider $\bm{\mathsf{Q}}(\bm{u}^*)$, the set of $\mathsf{Q}_i$ meeting at $\bm{u}^*\in\mathfrak{M}_\text{sing}^*$. Here, $\mathfrak{M}_\text{sing}^*$ is the set of isolated points where at least $N$ linearly independent hyperplanes meet. In this paper, we shall focus on the cases where the singularities are non-degenerate (namely, the number of hyperplanes meeting at $\bm{u}^*$ is equal to the total rank $N$ of the gauge group), which condition can be satisfied by uplifting with enough equivariant parameters.

The covector $\eta\in\mathfrak{h}^*$ picks out the allowed sets of hyperplanes in the JK residue. This is given by the positivity condition:
\begin{equation}
	\eta\in\text{Cone}(\mathsf{Q}_{i_j}):=\left\{\sum_{j=1}^Na_j \mathsf{Q}_{i_j}\,\Bigg|\,a_j\geq0\right\} \;.
\end{equation}
We shall refer to the hyperplanes/poles satisfying this positivity condition as admissible hyperplanes/poles (and inadmissible otherwise). In particular, for cyclic chambers where the crystal structures are well-known (at least for those arising from toric singularities), the choice is given by $\eta=(1,1,\dots,1)$.

\paragraph{The Jeffrey-Kirwan residue} The JK residue \cite{jeffrey1995localization} (see also \cite{Witten:1992xu}) is defined by
\begin{equation}
	\text{JK-Res}\, \frac{\text{d}\mathsf{Q}_{i_1}(u)}{\mathsf{Q}_{i_1}(u)}\wedge\dots\wedge\frac{\text{d}\mathsf{Q}_{i_N}(u)}{\mathsf{Q}_{i_N}(u)}:=\begin{cases}
		\text{sgn}(\text{det}(\mathsf{Q}_{i_1},\dots,\mathsf{Q}_{i_N}))\;,&\eta\in\text{Cone}(\mathsf{Q}_{i_j})\;,\\
		0\;,&\text{otherwise} \;.
	\end{cases}
\end{equation}
This can be rewritten as
\begin{equation}
	\text{JK-Res}\, \frac{\text{d}u_1\wedge\dots\wedge\text{d}u_N}{\mathsf{Q}_{i_1}(u)\dots \mathsf{Q}_{i_N}(u)}=\begin{cases}
		\displaystyle\frac{1}{|\text{det}(\mathsf{Q}_{i_1},\dots,\mathsf{Q}_{i_N})|} \;, &\eta\in\text{Cone}(\mathsf{Q}_{i_j}) \;,\\
		0\;,&\text{otherwise} \;.
	\end{cases}
\end{equation}
There is also a constructive definition of the JK residue \cite{szenes2003toric}. Readers are referred to, for example, \cite{Benini:2013xpa,Hori:2014tda,Cordova:2014oxa,Hwang:2014uwa,Bao:2024ygr,Bao:2025hfu} for more details.

\addcontentsline{toc}{section}{References}
\bibliographystyle{ytamsalpha}
\bibliography{references}

@article{Bao:2025hfu,
    author = "Bao, Jiakang and Yamazaki, Masahito",
    title = "{Crystals and Double Quiver Algebras from Jeffrey-Kirwan Residues}",
    eprint = "2501.03365",
    archivePrefix = "arXiv",
    primaryClass = "hep-th",
    month = "1",
    year = "2025"
}

@article{Galakhov:2021xum,
    author = "Galakhov, Dmitry and Li, Wei and Yamazaki, Masahito",
    title = "{Shifted quiver Yangians and representations from BPS crystals}",
    eprint = "2106.01230",
    archivePrefix = "arXiv",
    primaryClass = "hep-th",
    doi = "10.1007/JHEP08(2021)146",
    journal = "JHEP",
    volume = "08",
    pages = "146",
    year = "2021"
}

@article{Grekov:2023psy,
    author = "Grekov, Andrey and Nekrasov, Nikita",
    title = "{Elliptic analogue of the Vershik-Kerov limit shape.}",
    eprint = "2403.07168",
    archivePrefix = "arXiv",
    primaryClass = "math-ph",
    doi = "10.1134/S0016266324020059",
    journal = "Funct. Anal. Appl.",
    volume = "58",
    number = "2",
    pages = "143--159",
    year = "2023"
}

@article{Kimura:2023bxy,
    author = "Kimura, Taro and Noshita, Go",
    title = "{Gauge origami and quiver W-algebras}",
    eprint = "2310.08545",
    archivePrefix = "arXiv",
    primaryClass = "hep-th",
    doi = "10.1007/JHEP05(2024)208",
    journal = "JHEP",
    volume = "05",
    pages = "208",
    year = "2024"
}

@article{Nekrasov:2003rj,
    author = "Nekrasov, Nikita and Okounkov, Andrei",
    title = "{Seiberg-Witten theory and random partitions}",
    eprint = "hep-th/0306238",
    archivePrefix = "arXiv",
    reportNumber = "ITEP-TH-36-03, PUDM-2003, IHES-P-03-43",
    doi = "10.1007/0-8176-4467-9_15",
    journal = "Prog. Math.",
    volume = "244",
    pages = "525--596",
    year = "2006"
}

@article{Nekrasov:2002qd,
    author = "Nekrasov, Nikita A.",
    title = "{Seiberg-Witten prepotential from instanton counting}",
    eprint = "hep-th/0206161",
    archivePrefix = "arXiv",
    reportNumber = "ITEP-TH-22-02, IHES-P-04-22",
    doi = "10.4310/ATMP.2003.v7.n5.a4",
    journal = "Adv. Theor. Math. Phys.",
    volume = "7",
    number = "5",
    pages = "831--864",
    year = "2003"
}

@inproceedings{Losev:2003py,
    author = "Losev, Andrei S. and Marshakov, Andrei and Nekrasov, Nikita A.",
    title = "{Small instantons, little strings and free fermions}",
    booktitle = "{From Fields to Strings: Circumnavigating Theoretical Physics: A Conference in Tribute to Ian Kogan}",
    eprint = "hep-th/0302191",
    archivePrefix = "arXiv",
    reportNumber = "ITEP-TH-1-03, MPIM-2003-26, RN-FIAN-TD-05-03, IHES-P-03-09",
    pages = "581--621",
    month = "2",
    year = "2003"
}

@article{Nakajima:2003pg,
    author = "Nakajima, Hiraku and Yoshioka, Kota",
    title = "{Instanton counting on blowup. 1.}",
    eprint = "math/0306198",
    archivePrefix = "arXiv",
    doi = "10.1007/s00222-005-0444-1",
    journal = "Invent. Math.",
    volume = "162",
    pages = "313--355",
    year = "2005"
}

@article{Jiang:2025luj,
    author = "Jiang, Jiaqun",
    title = "{Shell formulas for instantons and gauge origami}",
    eprint = "2512.21606",
    archivePrefix = "arXiv",
    primaryClass = "hep-th",
    month = "12",
    year = "2025"
}

@article{Ooguri:2009ri,
    author = "Ooguri, Hirosi and Yamazaki, Masahito",
    title = "{Emergent Calabi-Yau Geometry}",
    eprint = "0902.3996",
    archivePrefix = "arXiv",
    primaryClass = "hep-th",
    reportNumber = "CALT-68-2721, IPMU09-0025, UT-09-04",
    doi = "10.1103/PhysRevLett.102.161601",
    journal = "Phys. Rev. Lett.",
    volume = "102",
    pages = "161601",
    year = "2009"
}

@article{Kenyon:2003uj,
    author = "Kenyon, Richard and Okounkov, Andrei and Sheffield, Scott",
    title = "{Dimers and amoebae}",
    eprint = "math-ph/0311005",
    archivePrefix = "arXiv",
    month = "11",
    year = "2003"
}

@article{cohn2001variational,
  title="{A variational principle for domino tilings}",
  author={Cohn, Henry and Kenyon, Richard and Propp, James},
  eprint = "math/0008220",
  archivePrefix = "arXiv",
  journal={Journal of the American Mathematical Society},
  volume={14},
  number={2},
  pages={297--346},
  year={2001}
}

@article{Franco:2015tna,
    author = "Franco, Sebastian and Ghim, Dongwook and Lee, Sangmin and Seong, Rak-Kyeong and Yokoyama, Daisuke",
    title = "{2d (0,2) Quiver Gauge Theories and D-Branes}",
    eprint = "1506.03818",
    archivePrefix = "arXiv",
    primaryClass = "hep-th",
    reportNumber = "CCNY-HEP-15-03, SNUTP15-003, KIAS-P15026, KCL-MTH-15-04",
    doi = "10.1007/JHEP09(2015)072",
    journal = "JHEP",
    volume = "09",
    pages = "072",
    year = "2015"
}

@article{Franco:2015tya,
    author = "Franco, Sebastian and Lee, Sangmin and Seong, Rak-Kyeong",
    title = "{Brane Brick Models, Toric Calabi-Yau 4-Folds and 2d (0,2) Quivers}",
    eprint = "1510.01744",
    archivePrefix = "arXiv",
    primaryClass = "hep-th",
    reportNumber = "CCNY-HEP-15-07, SNUTP15-009, KIAS-P15039",
    doi = "10.1007/JHEP02(2016)047",
    journal = "JHEP",
    volume = "02",
    pages = "047",
    year = "2016"
}

@article{Franco:2016nwv,
    author = "Franco, Sebastian and Lee, Sangmin and Seong, Rak-Kyeong",
    title = "{Brane brick models and 2d (0, 2) triality}",
    eprint = "1602.01834",
    archivePrefix = "arXiv",
    primaryClass = "hep-th",
    reportNumber = "CCNY-HEP-16-01, SNUTP-15-012, KIAS-P15062",
    doi = "10.1007/JHEP05(2016)020",
    journal = "JHEP",
    volume = "05",
    pages = "020",
    year = "2016"
}

@article{Franco:2016qxh,
    author = "Franco, Sebastian and Lee, Sangmin and Seong, Rak-Kyeong and Vafa, Cumrun",
    title = "{Brane Brick Models in the Mirror}",
    eprint = "1609.01723",
    archivePrefix = "arXiv",
    primaryClass = "hep-th",
    reportNumber = "CCNY-HEP-16-07, SNUTP16-004, KIAS-P16062",
    doi = "10.1007/JHEP02(2017)106",
    journal = "JHEP",
    volume = "02",
    pages = "106",
    year = "2017"
}

@article{Franco:2017cjj,
    author = "Franco, Sebastian and Ghim, Dongwook and Lee, Sangmin and Seong, Rak-Kyeong",
    title = "{Elliptic Genera of 2d (0,2) Gauge Theories from Brane Brick Models}",
    eprint = "1702.02948",
    archivePrefix = "arXiv",
    primaryClass = "hep-th",
    reportNumber = "CCNY-HEP-17-01, SNUTP17-002, UUITP-05-17",
    doi = "10.1007/JHEP06(2017)068",
    journal = "JHEP",
    volume = "06",
    pages = "068",
    year = "2017"
}

@article{Franco:2022iap,
    author = "Franco, Sebasti{\'a}n",
    title = "{2d Supersymmetric Gauge Theories, D-branes and Trialities}",
    eprint = "2201.10987",
    archivePrefix = "arXiv",
    primaryClass = "hep-th",
    month = "1",
    year = "2022"
}

@article{Franco:2022gvl,
    author = "Franco, Sebastian and Seong, Rak-Kyeong",
    title = "{Fano 3-folds, reflexive polytopes and brane brick models}",
    eprint = "2203.15816",
    archivePrefix = "arXiv",
    primaryClass = "hep-th",
    reportNumber = "UNIST-MTH-22-RS-01",
    doi = "10.1007/JHEP08(2022)008",
    journal = "JHEP",
    volume = "08",
    pages = "008",
    year = "2022"
}

@article{Franco:2022isw,
    author = "Franco, Sebastian and Ghim, Dongwook and Seong, Rak-Kyeong",
    title = "{Brane brick models for the Sasaki-Einstein 7-manifolds Y$^{p,k}$({\ensuremath{\mathbb{C}}}{\ensuremath{\mathbb{P}}}$^{1}${\texttimes} {\ensuremath{\mathbb{C}}}{\ensuremath{\mathbb{P}}}$^{1}$) and Y$^{p,k}$({\ensuremath{\mathbb{C}}}{\ensuremath{\mathbb{P}}}$^{2}$)}",
    eprint = "2212.02523",
    archivePrefix = "arXiv",
    primaryClass = "hep-th",
    reportNumber = "UNIST-MTH-22-RS-02, YITP-22-139",
    doi = "10.1007/JHEP03(2023)050",
    journal = "JHEP",
    volume = "03",
    pages = "050",
    year = "2023"
}

@article{Franco:2024lxs,
    author = "Franco, Sebasti{\'a}n",
    title = "{2d (0, 2) gauge theories from branes: Recent progress in brane brick models}",
    eprint = "2402.06993",
    archivePrefix = "arXiv",
    primaryClass = "hep-th",
    doi = "10.1142/S0217751X24460059",
    journal = "Int. J. Mod. Phys. A",
    volume = "39",
    number = "33",
    pages = "2446005",
    year = "2024"
}

@article{Okounkov:2003sp,
    author = "Okounkov, Andrei and Reshetikhin, Nikolai and Vafa, Cumrun",
    title = "{Quantum Calabi-Yau and classical crystals}",
    eprint = "hep-th/0309208",
    archivePrefix = "arXiv",
    reportNumber = "HUTP-03-A061",
    doi = "10.1007/0-8176-4467-9_16",
    journal = "Prog. Math.",
    volume = "244",
    pages = "597",
    year = "2006"
}

@article{Feng:2000mi,
    author = "Feng, Bo and Hanany, Amihay and He, Yang-Hui",
    title = "{D-brane gauge theories from toric singularities and toric duality}",
    eprint = "hep-th/0003085",
    archivePrefix = "arXiv",
    reportNumber = "MIT-CTP-2952",
    doi = "10.1016/S0550-3213(00)00699-4",
    journal = "Nucl. Phys. B",
    volume = "595",
    pages = "165--200",
    year = "2001"
}

@article{maillot1997g,
  title="{G\'eom\'etrie d'Arakelov des vari\'et\'es toriques et fibr\'es en droites int\'egrables}",
  author={Maillot, VINCENT},
  eprint = "alg-geom/9706005",
  archivePrefix = "arXiv",
  year={1997}
}

@article{roy2024generalized,
  title="{Generalized Mahler measures of Laurent polynomials}",
  author={Roy, Subham},
  eprint = "2308.04601",
  archivePrefix = "arXiv",
  primaryClass = "math.NT",
  journal={The Ramanujan Journal},
  volume={64},
  number={3},
  pages={581--627},
  year={2024},
  publisher={Springer}
}

@article{forsberg2000laurent,
  title="{Laurent determinants and arrangements of hyperplane amoebas}",
  author={Forsberg, Mikael and Passare, Mikael and Tsikh, August},
  journal={Advances in mathematics},
  volume={151},
  number={1},
  pages={45--70},
  year={2000},
  publisher={Elsevier}
}

@article{mikhalkin2001amoebas,
  title="{Amoebas of algebraic varieties}",
  author={Mikhalkin, Grigory},
  eprint = "math/0108225",
  archivePrefix = "arXiv",
  year={2001}
}

@article{Nekrasov:2015wsu,
    author = "Nekrasov, Nikita",
    title = "{BPS/CFT correspondence: non-perturbative Dyson-Schwinger equations and qq-characters}",
    eprint = "1512.05388",
    archivePrefix = "arXiv",
    primaryClass = "hep-th",
    doi = "10.1007/JHEP03(2016)181",
    journal = "JHEP",
    volume = "03",
    pages = "181",
    year = "2016"
}

@article{Kimura:2015rgi,
    author = "Kimura, Taro and Pestun, Vasily",
    title = "{Quiver W-algebras}",
    eprint = "1512.08533",
    archivePrefix = "arXiv",
    primaryClass = "hep-th",
    doi = "10.1007/s11005-018-1072-1",
    journal = "Lett. Math. Phys.",
    volume = "108",
    number = "6",
    pages = "1351--1381",
    year = "2018"
}

@article{Bao:2024ygr,
    author = "Bao, Jiakang and Seong, Rak-Kyeong and Yamazaki, Masahito",
    title = "{The origin of Calabi-Yau crystals in BPS states counting}",
    eprint = "2401.02792",
    archivePrefix = "arXiv",
    primaryClass = "hep-th",
    doi = "10.1007/JHEP03(2024)140",
    journal = "JHEP",
    volume = "03",
    pages = "140",
    year = "2024"
}

@article{Franco:2023tly,
    author = "Franco, Sebasti{\'a}n",
    title = "{4d crystal melting, toric Calabi-Yau 4-folds and brane brick models}",
    eprint = "2311.04404",
    archivePrefix = "arXiv",
    primaryClass = "hep-th",
    doi = "10.1007/JHEP03(2024)091",
    journal = "JHEP",
    volume = "03",
    pages = "091",
    year = "2024"
}

@article{Benini:2013xpa,
    author = "Benini, Francesco and Eager, Richard and Hori, Kentaro and Tachikawa, Yuji",
    title = "{Elliptic Genera of 2d ${\mathcal{N}}$ = 2 Gauge Theories}",
    eprint = "1308.4896",
    archivePrefix = "arXiv",
    primaryClass = "hep-th",
    reportNumber = "IPMU-13-0146, UT-13-29",
    doi = "10.1007/s00220-014-2210-y",
    journal = "Commun. Math. Phys.",
    volume = "333",
    number = "3",
    pages = "1241--1286",
    year = "2015"
}

@article{Hori:2014tda,
    author = "Hori, Kentaro and Kim, Heeyeon and Yi, Piljin",
    title = "{Witten Index and Wall Crossing}",
    eprint = "1407.2567",
    archivePrefix = "arXiv",
    primaryClass = "hep-th",
    reportNumber = "KIAS-P14039",
    doi = "10.1007/JHEP01(2015)124",
    journal = "JHEP",
    volume = "01",
    pages = "124",
    year = "2015"
}

@article{Cordova:2014oxa,
    author = "Cordova, Clay and Shao, Shu-Heng",
    title = "{An Index Formula for Supersymmetric Quantum Mechanics}",
    eprint = "1406.7853",
    archivePrefix = "arXiv",
    primaryClass = "hep-th",
    doi = "10.5427/jsing.2016.15b",
    journal = "J. Singul.",
    volume = "15",
    pages = "14--35",
    year = "2016"
}

@article{Hwang:2014uwa,
    author = "Hwang, Chiung and Kim, Joonho and Kim, Seok and Park, Jaemo",
    title = "{General instanton counting and 5d SCFT}",
    eprint = "1406.6793",
    archivePrefix = "arXiv",
    primaryClass = "hep-th",
    reportNumber = "SNUTP14-006",
    doi = "10.1007/JHEP07(2015)063",
    journal = "JHEP",
    volume = "07",
    pages = "063",
    year = "2015",
    note = "[Addendum: JHEP 04, 094 (2016)]"
}

@article{jeffrey1995localization,
  title="{Localization for nonabelian group actions}",
  author={Jeffrey, Lisa C and Kirwan, Frances C},
  eprint = "alg-geom/9307001",
  archivePrefix = "arXiv",
  journal={Topology},
  volume={34},
  number={2},
  pages={291--327},
  year={1995},
  publisher={Elsevier}
}

@article{Witten:1992xu,
    author = "Witten, Edward",
    title = "{Two-dimensional gauge theories revisited}",
    eprint = "hep-th/9204083",
    archivePrefix = "arXiv",
    doi = "10.1016/0393-0440(92)90034-X",
    journal = "J. Geom. Phys.",
    volume = "9",
    pages = "303--368",
    year = "1992"
}

@article{szenes2003toric,
  title="{Toric reduction and a conjecture of Batyrev and Materov}",
  author={Szenes, Andr{\'a}s and Vergne, Mich{\`e}le},
  eprint = "math/0306311",
  archivePrefix = "arXiv",
  year={2003}
}

@article{Benini:2013nda,
    author = "Benini, Francesco and Eager, Richard and Hori, Kentaro and Tachikawa, Yuji",
    title = "{Elliptic genera of two-dimensional N=2 gauge theories with rank-one gauge groups}",
    eprint = "1305.0533",
    archivePrefix = "arXiv",
    primaryClass = "hep-th",
    reportNumber = "IPMU-13-0082, UT-13-17",
    doi = "10.1007/s11005-013-0673-y",
    journal = "Lett. Math. Phys.",
    volume = "104",
    pages = "465--493",
    year = "2014"
}

@article{Witten:1986bf,
    author = "Witten, Edward",
    title = "{Elliptic Genera and Quantum Field Theory}",
    reportNumber = "PUPT-1024",
    doi = "10.1007/BF01208956",
    journal = "Commun. Math. Phys.",
    volume = "109",
    pages = "525",
    year = "1987"
}

@article{Witten:1982df,
    author = "Witten, Edward",
    title = "{Constraints on Supersymmetry Breaking}",
    reportNumber = "PRINT-82-0163 (PRINCETON)",
    doi = "10.1016/0550-3213(82)90071-2",
    journal = "Nucl. Phys. B",
    volume = "202",
    pages = "253",
    year = "1982"
}

@article{Bao:2025xhl,
    author = "Bao, Jiakang and Yamazaki, Masahito and Zhou, Dongao",
    title = "{Elliptic Genera of 2d $\mathcal{N}=(0,1)$ Gauge Theories}",
    eprint = "2508.06865",
    archivePrefix = "arXiv",
    primaryClass = "hep-th",
    month = "8",
    year = "2025"
}

@article{Ooguri:2009ijd,
    author = "Ooguri, Hirosi and Yamazaki, Masahito",
    title = "{Crystal Melting and Toric Calabi-Yau Manifolds}",
    eprint = "0811.2801",
    archivePrefix = "arXiv",
    primaryClass = "hep-th",
    reportNumber = "CALT-68-2706, IPMU-08-0087, UT-08-30",
    doi = "10.1007/s00220-009-0836-y",
    journal = "Commun. Math. Phys.",
    volume = "292",
    pages = "179--199",
    year = "2009"
}

@article{Szendroi:2007nu,
    author = "Szendroi, Balazs",
    title = "{Non-commutative Donaldson{\textendash}Thomas invariants and the conifold}",
    eprint = "0705.3419",
    archivePrefix = "arXiv",
    primaryClass = "math.AG",
    doi = "10.2140/gt.2008.12.1171",
    journal = "Geom. Topol.",
    volume = "12",
    number = "2",
    pages = "1171--1202",
    year = "2008"
}

@article{Chuang:2009crq,
    author = "Chuang, Wu-yen and Jafferis, Daniel L.",
    title = "{Wall Crossing of BPS States on the Conifold from Seiberg Duality and Pyramid Partitions}",
    eprint = "0810.5072",
    archivePrefix = "arXiv",
    primaryClass = "hep-th",
    reportNumber = "RUNHETC-2008-11",
    doi = "10.1007/s00220-009-0832-2",
    journal = "Commun. Math. Phys.",
    volume = "292",
    pages = "285--301",
    year = "2009"
}

@article{Dimofte:2009bv,
    author = "Dimofte, Tudor and Gukov, Sergei",
    title = "{Refined, Motivic, and Quantum}",
    eprint = "0904.1420",
    archivePrefix = "arXiv",
    primaryClass = "hep-th",
    reportNumber = "CALT-68-2725",
    doi = "10.1007/s11005-009-0357-9",
    journal = "Lett. Math. Phys.",
    volume = "91",
    pages = "1",
    year = "2010"
}

@article{Nagao:2009rq,
    author = "Nagao, Kentaro and Yamazaki, Masahito",
    title = "{The Non-commutative Topological Vertex and Wall Crossing Phenomena}",
    eprint = "0910.5479",
    archivePrefix = "arXiv",
    primaryClass = "hep-th",
    reportNumber = "CALT-68-2755, IPMU09-0132, UT-09-24",
    doi = "10.4310/ATMP.2010.v14.n4.a3",
    journal = "Adv. Theor. Math. Phys.",
    volume = "14",
    number = "4",
    pages = "1147--1181",
    year = "2010"
}

@article{Aganagic:2009cg,
    author = "Aganagic, Mina and Yamazaki, Masahito",
    title = "{Open BPS Wall Crossing and M-theory}",
    eprint = "0911.5342",
    archivePrefix = "arXiv",
    primaryClass = "hep-th",
    reportNumber = "CALT-68-2763, IPMU09-0142, UT-09-26",
    doi = "10.1016/j.nuclphysb.2010.03.019",
    journal = "Nucl. Phys. B",
    volume = "834",
    pages = "258--272",
    year = "2010"
}

@article{Ooguri:2010yk,
    author = "Ooguri, Hirosi and Sulkowski, Piotr and Yamazaki, Masahito",
    title = "{Wall Crossing As Seen By Matrix Models}",
    eprint = "1005.1293",
    archivePrefix = "arXiv",
    primaryClass = "hep-th",
    reportNumber = "AEI-2010-091, CALT-68-2786, IPMU10-0078",
    doi = "10.1007/s00220-011-1330-x",
    journal = "Commun. Math. Phys.",
    volume = "307",
    pages = "429--462",
    year = "2011"
}

@article{Cirafici:2010bd,
    author = "Cirafici, Michele and Sinkovics, Annamaria and Szabo, Richard J.",
    title = "{Instantons, Quivers and Noncommutative Donaldson-Thomas Theory}",
    eprint = "1012.2725",
    archivePrefix = "arXiv",
    primaryClass = "hep-th",
    reportNumber = "DAMTP-2010-124, HWM-10-35, EMPG-10-25, DAMTP-2010-124-, HWM-10-35-",
    doi = "10.1016/j.nuclphysb.2011.08.002",
    journal = "Nucl. Phys. B",
    volume = "853",
    pages = "508--605",
    year = "2011"
}

@article{Yamazaki:2010fz,
    author = "Yamazaki, Masahito",
    title = "{Crystal Melting and Wall Crossing Phenomena}",
    eprint = "1002.1709",
    archivePrefix = "arXiv",
    primaryClass = "hep-th",
    reportNumber = "CALT-68-2773, IPMU10-0023, UT-10-01",
    doi = "10.1142/S0217751X11051482",
    journal = "Int. J. Mod. Phys. A",
    volume = "26",
    pages = "1097--1228",
    year = "2011"
}

@article{Yamazaki:2011wy,
    author = "Yamazaki, Masahito",
    title = "{Geometry and Combinatorics of Crystal Melting}",
    eprint = "1102.0776",
    archivePrefix = "arXiv",
    primaryClass = "math-ph",
    reportNumber = "PUPT-2365",
    journal = "RIMS Kokyuroku Bessatsu B",
    volume = "28",
    pages = "193",
    year = "2011"
}

@article{Nekrasov:2017cih,
    author = "Nekrasov, Nikita",
    title = "{Magnificent four}",
    eprint = "1712.08128",
    archivePrefix = "arXiv",
    primaryClass = "hep-th",
    doi = "10.4310/ATMP.2020.v24.n5.a4",
    journal = "Adv. Theor. Math. Phys.",
    volume = "24",
    number = "5",
    pages = "1171--1202",
    year = "2020"
}

@article{Nekrasov:2018xsb,
    author = "Nekrasov, Nikita and Piazzalunga, Nicol{\`o}",
    title = "{Magnificent Four with Colors}",
    eprint = "1808.05206",
    archivePrefix = "arXiv",
    primaryClass = "hep-th",
    doi = "10.1007/s00220-019-03426-3",
    journal = "Commun. Math. Phys.",
    volume = "372",
    number = "2",
    pages = "573--597",
    year = "2019"
}

@article{Nekrasov:2023nai,
    author = "Nekrasov, Nikita and Piazzalunga, Nicol{\`o}",
    title = "{Global Magni4icence, or: 4G Networks}",
    eprint = "2306.12995",
    archivePrefix = "arXiv",
    primaryClass = "hep-th",
    doi = "10.3842/SIGMA.2024.106",
    journal = "SIGMA",
    volume = "20",
    pages = "106",
    year = "2024"
}

@article{Galakhov:2023vic,
    author = "Galakhov, Dmitry and Li, Wei",
    title = "{Charging solid partitions}",
    eprint = "2311.02751",
    archivePrefix = "arXiv",
    primaryClass = "hep-th",
    doi = "10.1007/JHEP01(2024)043",
    journal = "JHEP",
    volume = "01",
    pages = "043",
    year = "2024"
}

@article{Szabo:2024lcp,
    author = "Szabo, Richard J. and Tirelli, Michelangelo",
    title = "{Tetrahedron instantons on orbifolds}",
    eprint = "2405.14792",
    archivePrefix = "arXiv",
    primaryClass = "hep-th",
    doi = "10.1007/s11005-025-01903-6",
    journal = "Lett. Math. Phys.",
    volume = "115",
    number = "1",
    pages = "11",
    year = "2025"
}

@article{Bao:2025dqs,
    author = "Bao, Jiakang",
    title = "{An overview of crystals and double quiver Yangians}",
    eprint = "2509.16918",
    archivePrefix = "arXiv",
    primaryClass = "hep-th",
    doi = "10.1142/S0217751X25300182",
    journal = "Int. J. Mod. Phys. A",
    volume = "41",
    number = "01",
    pages = "2530018",
    year = "2026"
}

@article{Li:2020rij,
    author = "Li, Wei and Yamazaki, Masahito",
    title = "{Quiver Yangian from Crystal Melting}",
    eprint = "2003.08909",
    archivePrefix = "arXiv",
    primaryClass = "hep-th",
    reportNumber = "IPMU-20-0027",
    doi = "10.1007/JHEP11(2020)035",
    journal = "JHEP",
    volume = "11",
    pages = "035",
    year = "2020"
}

@article{Galakhov:2020vyb,
    author = "Galakhov, Dmitry and Yamazaki, Masahito",
    title = "{Quiver Yangian and Supersymmetric Quantum Mechanics}",
    eprint = "2008.07006",
    archivePrefix = "arXiv",
    primaryClass = "hep-th",
    doi = "10.1007/s00220-022-04490-y",
    journal = "Commun. Math. Phys.",
    volume = "396",
    number = "2",
    pages = "713--785",
    year = "2022"
}

@article{Galakhov:2021vbo,
    author = "Galakhov, Dmitry and Li, Wei and Yamazaki, Masahito",
    title = "{Toroidal and elliptic quiver BPS algebras and beyond}",
    eprint = "2108.10286",
    archivePrefix = "arXiv",
    primaryClass = "hep-th",
    doi = "10.1007/JHEP02(2022)024",
    journal = "JHEP",
    volume = "02",
    pages = "024",
    year = "2022"
}

@article{Noshita:2021ldl,
    author = "Noshita, Go and Watanabe, Akimi",
    title = "{A note on quiver quantum toroidal algebra}",
    eprint = "2108.07104",
    archivePrefix = "arXiv",
    primaryClass = "hep-th",
    doi = "10.1007/JHEP05(2022)011",
    journal = "JHEP",
    volume = "05",
    pages = "011",
    year = "2022"
}

@article{Yamazaki:2022cdg,
    author = "Yamazaki, Masahito",
    title = "{Quiver Yangians and crystal meltings: A concise summary}",
    eprint = "2203.14314",
    archivePrefix = "arXiv",
    primaryClass = "hep-th",
    doi = "10.1063/5.0089785",
    journal = "J. Math. Phys.",
    volume = "64",
    number = "1",
    pages = "011101",
    year = "2023"
}

@article{Kimura:2016dys,
    author = "Kimura, Taro and Pestun, Vasily",
    title = "{Quiver elliptic W-algebras}",
    eprint = "1608.04651",
    archivePrefix = "arXiv",
    primaryClass = "hep-th",
    doi = "10.1007/s11005-018-1073-0",
    journal = "Lett. Math. Phys.",
    volume = "108",
    number = "6",
    pages = "1383--1405",
    year = "2018"
}

@article{Koroteev:2019byp,
    author = "Koroteev, Peter",
    title = "{On Quiver W-algebras and Defects from Gauge Origami}",
    eprint = "1908.04394",
    archivePrefix = "arXiv",
    primaryClass = "hep-th",
    doi = "10.1016/j.physletb.2019.135101",
    journal = "Phys. Lett. B",
    volume = "800",
    pages = "135101",
    year = "2020"
}

@article{Kimura:2024xpr,
    author = "Kimura, Taro and Noshita, Go",
    title = "{Gauge origami and quiver W-algebras II: Vertex function and beyond quantum $q$-Langlands correspondence}",
    eprint = "2404.17061",
    archivePrefix = "arXiv",
    primaryClass = "hep-th",
    month = "4",
    year = "2024"
}

@article{Kimura:2024osv,
    author = "Kimura, Taro and Noshita, Go",
    title = "{Gauge origami and quiver W-algebras. Part III. Donaldson-Thomas qq-characters}",
    eprint = "2411.01987",
    archivePrefix = "arXiv",
    primaryClass = "hep-th",
    doi = "10.1007/JHEP03(2025)050",
    journal = "JHEP",
    volume = "03",
    pages = "050",
    year = "2025"
}

@article{Kimura:2025lfo,
    author = "Kimura, Taro and Noshita, Go",
    title = "{Gauge origami and quiver W-algebras. Part IV. Pandharipande-Thomas qq-characters}",
    eprint = "2508.12125",
    archivePrefix = "arXiv",
    primaryClass = "hep-th",
    reportNumber = "TIT/HEP-706",
    doi = "10.1007/JHEP01(2026)063",
    journal = "JHEP",
    volume = "01",
    pages = "063",
    year = "2026"
}

@article{Kimura:2025lig,
    author = "Kimura, Taro and Noshita, Go",
    title = "{The 4-fold Pandharipande{\textendash}Thomas vertex and Jeffrey{\textendash}Kirwan residue}",
    eprint = "2508.12128",
    archivePrefix = "arXiv",
    primaryClass = "hep-th",
    doi = "10.1007/s11005-026-02071-x",
    journal = "Lett. Math. Phys.",
    volume = "116",
    number = "2",
    pages = "45",
    year = "2026"
}

@article{Nekrasov:2016ydq,
    author = "Nekrasov, Nikita",
    title = "{BPS/CFT Correspondence III: Gauge Origami partition function and qq-characters}",
    eprint = "1701.00189",
    archivePrefix = "arXiv",
    primaryClass = "hep-th",
    doi = "10.1007/s00220-017-3057-9",
    journal = "Commun. Math. Phys.",
    volume = "358",
    number = "3",
    pages = "863--894",
    year = "2018"
}

@article{Carcamo:2026yqu,
    author = "Carcamo, Mario and Franco, Sebasti{\'a}n",
    title = "{Crystal Melting, Triality and Partition Functions for Toric Calabi-Yau Fourfolds}",
    eprint = "2603.08815",
    archivePrefix = "arXiv",
    primaryClass = "hep-th",
    month = "3",
    year = "2026"
}

@article{Iqbal:2003ds,
    author = "Iqbal, Amer and Nekrasov, Nikita and Okounkov, Andrei and Vafa, Cumrun",
    title = "{Quantum foam and topological strings}",
    eprint = "hep-th/0312022",
    archivePrefix = "arXiv",
    reportNumber = "HUTP-03-A078, IHES-P-03-65, ITEP-TH-60-03",
    doi = "10.1088/1126-6708/2008/04/011",
    journal = "JHEP",
    volume = "04",
    pages = "011",
    year = "2008"
}

@article{Galakhov:2024foa,
    author = "Galakhov, Dmitry and Morozov, Alexei and Tselousov, Nikita",
    title = "{Wall-crossing effects on quiver BPS algebras}",
    eprint = "2403.14600",
    archivePrefix = "arXiv",
    primaryClass = "hep-th",
    doi = "10.1007/JHEP05(2024)118",
    journal = "JHEP",
    volume = "05",
    pages = "118",
    year = "2024"
}

@article{Yamazaki:2012cp,
    author = "Yamazaki, Masahito",
    title = "{Quivers, YBE and 3-manifolds}",
    eprint = "1203.5784",
    archivePrefix = "arXiv",
    primaryClass = "hep-th",
    reportNumber = "PUPT-2406",
    doi = "10.1007/JHEP05(2012)147",
    journal = "JHEP",
    volume = "05",
    pages = "147",
    year = "2012"
}

@incollection {Ronkin,
    AUTHOR = {Ronkin, L. I.},
     TITLE = {On zeros of almost periodic functions generated by functions
              holomorphic in a multicircular domain},
 BOOKTITLE = {Complex analysis in modern mathematics (Russian)},
     PAGES = {239--251},
 PUBLISHER = {FAZIS, Moscow},
      YEAR = {2001},
   MRCLASS = {32A07 (32A22)},
  MRNUMBER = {MR1833516 (2002c:32003)},
MRREVIEWER = {Serguey V. Shvedenko},
}

@article {PassareRullgard,
    AUTHOR = {Passare, Mikael and Rullg{\aa}rd, Hans},
     TITLE = {Amoebas, {M}onge-{A}mp\`ere measures, and triangulations of
              the {N}ewton polytope},
   JOURNAL = {Duke Math. J.},
  FJOURNAL = {Duke Mathematical Journal},
    VOLUME = {121},
      YEAR = {2004},
    NUMBER = {3},
     PAGES = {481--507},
      ISSN = {0012-7094},
     CODEN = {DUMJAO},
   MRCLASS = {32A60 (52A41 52B20)},
  MRNUMBER = {MR2040284 (2005a:32005)},
MRREVIEWER = {A. Yu. Rashkovski{\u\i}},
}

@Article{Kasteleyn,
     author    = "P. W. Kasteleyn",
     title     = "The statistics of dimers on a lattice",
     journal   = "Physica",
     volume    = "27",
     year      = "1961",
     pages     = "1209"
}

@article{vandervelde2003formula,
  title="{A formula for the Mahler measure of $axy+ bx+ cy+ d$}",
  author={Vandervelde, Sam},
  journal={Journal of Number Theory},
  volume={100},
  number={1},
  pages={184--202},
  year={2003},
  publisher={Elsevier}
}

@article{Zahabi:2018znq,
    author = "Zahabi, Ali",
    title = "{Toric Quiver Asymptotics and Mahler Measure: $\mathcal{N}=2$ BPS States}",
    eprint = "1812.10287",
    archivePrefix = "arXiv",
    primaryClass = "hep-th",
    doi = "10.1007/JHEP07(2019)121",
    journal = "JHEP",
    volume = "07",
    pages = "121",
    year = "2019"
}

@article{de2021limit,
  title="{Limit shapes for the asymmetric five vertex model}",
  author={de Gier, Jan and Kenyon, Richard and Watson, Samuel S},
  eprint = "1812.11934",
  archivePrefix = "arXiv",
  primaryClass = "math.PR",
  journal={Communications in Mathematical Physics},
  volume={385},
  number={2},
  pages={793--836},
  year={2021},
  publisher={Springer}
}

@article{Pomoni:2023nlf,
    author = "Pomoni, Elli and Yan, Wenbin and Zhang, Xinyu",
    title = "{Probing M-theory with tetrahedron instantons}",
    eprint = "2306.06005",
    archivePrefix = "arXiv",
    primaryClass = "hep-th",
    reportNumber = "DESY-23-073",
    doi = "10.1007/JHEP11(2023)177",
    journal = "JHEP",
    volume = "11",
    pages = "177",
    year = "2023"
}

@article{Fasola:2023ypx,
    author = "Fasola, Nadir and Monavari, Sergej",
    title = "{Tetrahedron instantons in Donaldson-Thomas theory}",
    eprint = "2306.07145",
    archivePrefix = "arXiv",
    primaryClass = "math.AG",
    doi = "10.1016/j.aim.2024.110099",
    journal = "Adv. Math.",
    volume = "462",
    pages = "110099",
    year = "2025"
}

\end{document}